\documentclass[a4paper,fleqn,usenatbib]{mnras}

\usepackage{newtxtext,newtxmath}

\usepackage[T1]{fontenc}
\usepackage{ae,aecompl}

\usepackage{tikz}

\usepackage{graphicx}	
\usepackage{amsmath}	
\usepackage{amssymb}	

\title[Radio observations of Abell 1240]{Radio observations of the double-relic galaxy cluster Abell 1240}

\author[D. N. Hoang et al.]{D. N. Hoang$^{1}$\thanks{E-mail: hoang@strw.leidenuniv.nl},
T. W. Shimwell$^{2,1}$,
R. J. van Weeren$^{1}$,
H. T. Intema$^{1}$,\newauthor
H. J. A. R\"{o}ttgering$^{1}$,
F. Andrade-Santos$^{3}$,
H. Akamatsu$^{4}$,
A. Bonafede$^{5,6}$,\newauthor
G. Brunetti$^{5}$,
W. A. Dawson$^{7}$,
N. Golovich$^{8,7}$,
P. N. Best$^{9}$,
A. Botteon$^{10,5}$,\newauthor
M. Br\"uggen$^{6}$, 
R. Cassano$^{5}$,
F. de Gasperin$^{1}$,
M. Hoeft$^{11}$,
A. Stroe$^{12}$\thanks{ESO fellow} and \newauthor
G. J. White$^{13,14}$
\\
\textit{(Affiliations are listed at the end of the paper)}
}

\date{Accepted 2018. Received 2018; in original form 2018}

\pubyear{2017}

\begin{document}
\label{firstpage}
\pagerange{\pageref{firstpage}--\pageref{lastpage}}
\maketitle

\begin{abstract}
We present LOFAR $120-168$ MHz images of the merging galaxy cluster Abell 1240 that hosts double radio relics. In combination with the GMRT $595-629$ MHz and VLA $2-4$ GHz data, we characterised the spectral and polarimetric properties of the radio emission. The spectral indices for the relics steepen from their outer edges towards the cluster centre and the electric field vectors are approximately perpendicular to the major axes of the relics. The results are consistent with the picture that these relics trace large-scale shocks propagating outwards during the merger. Assuming diffusive shock acceleration (DSA), we obtain shock Mach numbers of $\mathcal{M}=2.4$ and $2.3$ for the northern and southern shocks, respectively. For $\mathcal{M}\lesssim3$ shocks, a pre-existing population of mildly relativistic electrons is required to explain the brightness of the relics due to the high ($>10$ per cent) particle acceleration efficiency required. However, for $\mathcal{M}\gtrsim4$ shocks the required efficiency is $\gtrsim1\%$ and $\gtrsim0.5\%$, respectively, which is low enough for shock acceleration directly from the thermal pool. We used the fractional polarization to constrain the viewing angle to $\geqslant53\pm3^\circ$ and $\geqslant39\pm5^\circ$ for the northern and southern shocks, respectively. We found no evidence for diffuse emission in the cluster central region. If the halo spans the entire region between the relics ($\sim1.8\,\text{Mpc}$) our upper limit on the power is $P_\text{\tiny 1.4 GHz}=(1.4\pm0.6)\times10^{23}\,\text{W\,Hz}^{-1}$ which is approximately equal to the anticipated flux from a cluster of this mass. However, if the halo is smaller than this, our constraints on the power imply that the halo is underluminous.
\end{abstract}

\begin{keywords}
galaxies: clusters: individual (Abell 1240)
\textendash{} galaxies: clusters: intra-cluster medium \textendash{}
large-scale structure of Universe \textendash{} radiation mechanisms:
non-thermal \textendash{} diffuse radiation \textendash{} shock waves
\end{keywords}


\section{Introduction}
\label{sec:intro}

Massive galaxy clusters consist of hundreds to thousands of galaxies and grow hierarchically through a sequence of mergers of smaller clusters or groups of galaxies. Merging events between massive clusters release an enormous amount of gravitational energy \mbox{($\sim10^{63}-10^{64}$ ergs)} to the intra-cluster medium (ICM) over a few Gyrs \citep[e.g.][]{Hoeft2008,Brunetti2014}. Most of this energy is transferred to thermal energy by heating of the ICM plasma. Through large-scale motions (i.e. shocks and turbulence) and magnetic field amplification, a small fraction of it is converted to non-thermal energy of relativistic particles that permeate the ICM. In the presence of the large-scale, $\sim\mu$G cluster magnetic field, these relativistic electrons, that have a Lorentz factor $\sim10^3-10^5$, emit synchrotron emission that is observable in the radio band \citep[see, e.g.,][]{Blandford1987,Luigina2012,Brunetti2014}.

Depending on the morphology, location and polarimetric properties, the diffuse radio sources in galaxy clusters are primarily classified as radio haloes or relics. Radio relics are elongated diffuse sources observed at the periphery of galaxy clusters; some relics have been detected with a high fraction of linear polarization at $\sim$ GHz frequencies (i.e. from $\sim10\%$ up to $70\%$) and distinctive spectral index gradients across their structure. Radio haloes are cluster-wide sources that roughly follow the X-ray emission and are observed to be unpolarized down to a few percent \citep[see, e.g.,][]{Luigina2012,Kierdorf2017}.

The formation mechanism of radio haloes and relics has not been fully understood. The prominent model for the generation of radio haloes is the turbulent re-acceleration model where relativistic electrons are re-energized by magnetohydrodynamical turbulence during cluster mergers \citep{Brunetti2001,Petrosian2001a}. Another model has been proposed to explain the existence of radio haloes such as the hadronic model in which relativistic electrons are \textit{secondary} products of proton-proton collisions \citep[e.g. ][]{Dennison1980a}. However, the \textit{secondary} model has difficulties because of the non-detection of $\gamma$-rays by  the \textit{Fermi $\gamma$-ray Space Telescope} \citep[e.g.][]{Jeltema2009,Ackermann2010,Jeltema2011,Brunetti2012,Zandanel2014,Ackermann2016b, Brunetti2017}. The \textit{secondary} model is further challenged by the large amount of energy that is required to explain the steep spectra of some radio haloes \citep[e.g.][]{Brunetti2004a,Brunetti2008}. However, the observed radio emission may be caused by a combination of the mechanisms in which the secondary electrons are re-accelerated by merger turbulence  \citep{Brunetti2005,Brunetti2011b,Pinzke2017}.

Radio relics have been proposed to be the synchrotron emission from large-scale shocks that are generated during cluster mergers \citep[e.g.][]{Ensslin1998}. Relativistic electrons gain energy either from the direct acceleration of the ICM thermal electrons or from the re-acceleration of fossil plasma remnants of active galactic nuclei (AGN) through Fermi-I DSA  \citep[e.g.][]{Giacintucci2008,Markevitch2005}. Observational evidence associating the formation of radio relics with cluster merger shocks have been observed in a large number of merging clusters (e.g. \citealt{VanWeeren2010a,Bonafede2012,Stroe2013a,DeGasperin2015,VanWeeren2016b} or see \citealt{Luigina2012} for a review). The evidence includes (i) an arc-like morphology of some relics, which is consistent with an edge-on/close to edge-on view of 3D shock waves, (ii) spectral gradients or spectral curvature variations across the width of relics, suggesting that the relativistic electrons gain energy at the shock fronts and lose their energy after shock passage and (iii) high degree of linear polarization, indicating a magnetic field aligned within the shock plane. The distribution of size, shape and location of relics agree well with those of merger shocks in cosmological simulations \citep[e.g.][]{Nuza2017}.  Alternatively, the re-acceleration model requires pre-existing populations of mildly relativistic electrons to be present in the regions of the shocks and there is evidence for this in a few cases \citep[e.g.][]{weeren2013,Bonafede2014,shimwell2015,Botteon2016a,VanWeeren2017}.
	
Galaxy clusters that host double radio relics on diametrically opposite sides of the clusters are some of the most interesting cases to study particle \mbox{(re-)acceleration} at Mpc scales. Only 17 double-relics clusters have been detected to date (\citealt{Bonafede2017} and references therein). In these rare energetic merging clusters both the relics and the halo are expected to be generated by shocks and turbulence, respectively. Due to the diametrically opposite locations of the double relics, these shocks are thought to be caused by head-on binary mergers of roughly equal masses merging on/close to the plane of the sky \citep[e.g.][]{VanWeeren2011,Nuza2017}. Hence, double-relic clusters provide a unique environment for studies of particle \mbox{(re-)acceleration} without the complication of projection effects (i.e. to minimise a mixture of relativistic electron populations along the line-of-sight (LOS); \citealt{Stroe2013a}). Furthermore, since the possibility to have seed populations of mildly relativistic electrons is likely to correlate with the distribution of aged AGN, these double relics might provide hints as to whether relics are formed by acceleration directly from the thermal pool or from fossil plasma pre-existing in the ICM. 

Abell 1240 ($z=0.1948$; hereafter A1240) is a binary merging galaxy cluster ($M_{500}=(3.7\pm0.4)\times10^{14}\,M_\odot$; \citealt{Planck2015}). A1240 was first observed to host faint diffuse emission located on the opposite sides of the cluster by \cite{Kempner2001}. Follow-up observations by \cite{Bonafede2009} confirmed the existence of the radio relics (labelled as A1240-1 and A1240-2 for the northern and southern relics, respectively) and found that they are elongated over $\sim650\,\text{kpc}$ and $\sim1250\,\text{kpc}$ in the east-west direction. Across the width of A1240-1, the spectral indices\footnote{The  convention $S\propto\nu^\alpha$  is used in this paper} steepen from $-1.1$ to $-1.6$ towards the cluster centre. Polarized emission was observed from the relics at 1.4 GHz and the electric field vectors are approximately perpendicular to the major axes of the relics, indicating the alignment of the ICM magnetic fields. Assuming a relativistic electron energy of Lorentz factor $\geq100$ with the spectrum of $N(p)\propto p^{-\delta}$ (where $\delta=-2\alpha_\text{\tiny int}$, $\alpha_\text{\tiny int}^{\text{\tiny A1240-1}}=-1.2$ and $\alpha_\text{\tiny int}^{\text{\tiny A1240-2}}=-1.3$) and equipartition energy conditions, \cite{Bonafede2009} estimated the equipartition magnetic field of $2.4\,\mu$G and $2.5\,\mu$G for A1240-1 and A1240-2, respectively. Due to their properties (i.e. location, morphology, spectral gradients and polarization properties), the relics were interpreted as synchrotron emission from large-scale shocks that were generated by a cluster merger in the plane of the sky and are moving outwards. Using the integrated spectral indices, \cite{Bonafede2009} estimated the Mach numbers of $\mathcal{M_\text{\tiny int}}=3.3\pm0.2$  for A1240-1 and $\mathcal{M_\text{\tiny int}}=2.8\pm0.3$ for A1240-2.

In the optical band, \cite{Barrena2009} studied the dynamical properties of A1240 using spectroscopic redshifts from 145 galaxies. A1240 was found to have a bimodal structure with clumps of galaxies separated in the north-south direction. The galaxy clumps have a relative rest-frame LOS velocity of $V_\text{\tiny rf}=390\,\text{km\,s}^{-1}$ at a projected distance of $1.2\,h^{-1}_{70}$ Mpc. The galaxy clumps were estimated to have passed $0.3$ Gyr ago. Approximately $12\arcmin$ ($\sim2.3\,\text{Mpc}$) to the south of A1240 is Abell 1237 (hereafter A1237; $z=0.1935$, \citealt{Barrena2009}) that is thought to be in-falling to A1240. \cite{Barrena2009} found no signature of peculiar displacement of A1240 towards the direction of A1237 and suggested that A1237 and A1240 are in the pre-merging stage.

In this paper, we present Low Frequency Array (LOFAR) $120-187$ MHz observations of A1240. LOFAR's sensitivity to large-scale emission coupled with its high-angular resolution helps us to study the extended diffuse emission from A1240 in detail. Furthermore, LOFAR observations at low frequencies $\leqslant200$ MHz allow us to detect steep spectrum emission such as from radio haloes that are generated during minor mergers or mergers of low-mass clusters. In combination with archival Giant Metrewave Radio Telescope (GMRT) $595-629$ MHz and Karl G. Jansky Very Large Array (VLA) $2-4$ GHz data, we study spatial variations of spectral indices of the radio sources in A1240 across a wide frequency range. We use the VLA data to study the polarized emission from the cluster relics.

Throughout this paper, we assume $H_{0}=70$ km s$^{-1}$  Mpc$^{-1}$, $\Omega_{M}=0.3$ and  $\Omega_{\Lambda}=0.7$. In this cosmology, $1\arcmin$ corresponds to $\sim194h_{70}^{-1}$ kpc at the cluster redshift of $z=0.1948$.

\section{Observations and data reduction}

\subsection{LOFAR 143 MHz}
\label{sec:red_lofar}

A1240 was observed with LOFAR for 8 hours on March 22, 2016 as part of the LOFAR Two-meter Sky Survey (LoTSS, \citealt{Shimwell2017}). A1240 was located at an angular distance of $\sim55\arcmin$ from the LoTSS grid pointing centre (pointing P170+42 of project LC4\_034), where the primary beam sensitivity is $\sim78$ percent of the value at the pointing centre. The observations used the High-Band Antennas ($120-187$ MHz) on 48 core, 14 remote and 9 international stations. However, to obtain cluster maps at spatial resolutions of $>8\arcsec$, we used only data from baselines that have $uv$-range between 15 $\lambda$ and 66 k$\lambda$. For calibration purposes, 3C 196 was observed for 10 minutes before the target field. Details of the observations are listed in Table \ref{tab:lofar_gmrt_vla_obs}.

\begin{table*}
	\centering
	\caption{LOFAR, GMRT and VLA observations}  
	\begin{tabular}{lccc}
		\hline\hline
		                       &      LOFAR 143 MHz      &      GMRT 612 MHz       &             VLA 3 GHz             \\ \hline
		Pointing (RA, Dec)     & $170^\text{h}48^\text{m}54.0^\text{s}$, $+42^\text{d}10^\text{m}13.08^\text{s}$ & $11^\text{h}23^\text{m}32.1^\text{s}$, $+43^\text{d}06^\text{m}31.9^\text{s}$ & $11^\text{h}23^\text{m}29.0^\text{s}$, $+43^\text{d}09^\text{m}42.0^\text{s}$ (A1240-1) \\
		                       &                         &                         & $11^\text{h}23^\text{m}35.0^\text{s}$, $+43^\text{d}01^\text{m}12.9^\text{s}$ (A1240-2) \\
		Configuration          &           N/A           &           N/A           &               C, D                \\
		Observation date       &     March 22, 2016      &      July 2, 2011       &         October 10, 2014 (C)         \\
		                       &                         &                         &         October 09, 2015 (D)         \\
		On-source time (hr)    &           8.0           &           6.0           &         1.1 (C), 3.4 (D)          \\
		Freq. coverage (GHz)   &       $0.120-0.187$       &       $0.595-0.629$       &                $2-4 $               \\
		Usable bandwidth (GHz) &          0.043          &         0.0333          &               1.992               \\
		Channel width (MHz)    &         0.0122          &          0.13           &                 2                 \\
		Integration time (s)   &            1            &           16            &              $\sim5$              \\
		Correlation            &     XX, XY, YX, YY      &         RR, LL          &          RR, RL, LR, LL           \\
		Number of antennas     &           62            &           28            &                51                 \\ \hline\hline
	\end{tabular}  
	\label{tab:lofar_gmrt_vla_obs}
\end{table*}

We used the facet calibration scheme to calibrate the target data for both direction-independent and direction-dependent effects. Details of the facet calibration are given in \cite{VanWeeren2016a,Williams2016a}. But for completeness, we briefly describe the procedure below.

During the direction-independent calibration part, the target data were flagged for radio frequency interference (RFI) using $\mathtt{AOFlagger}$ \citep{Offringa2012c} and time periods where the contamination from bright radio sources in the sidelobes (i.e. Cassiopeia A, Hercules A, Taurus A and Virgo A) were also flagged. The amplitude gains, initial clock and XX-YY phase offsets were derived from gain solutions of 3C 196. Here the calibration solutions of 3C 196 were obtained by calibrating a 4-Gaussian component source model that has an integrated flux of 83.1 Jy at 150 MHz in agreement with the \citet{Scaife2012} absolute flux scale.

To prepare for the direction-dependent calibration, \lq\lq{blank}\rq\rq \,  data sets for the target field were made by subtracting all sources within a $\sim30^\circ$ radius of the field centre using the direction independent calibration solutions. The $\mathtt{CLEAN}$ components used for the subtraction were obtained from imaging at resolutions of $\sim40\arcsec$ and $\sim2\arcmin$.

In the direction-dependent calibration part, we aimed to obtain thermal-noise limited images of the cluster. To achieve this, the ionospheric distortions and beam errors towards the target direction were corrected and the contamination from nearby sources was minimized following the facet calibration procedure. The target field was divided into 15 regions (called facets), each containing bright, compact calibrator source(s). The direction-dependent gain and TEC (total electron content) solutions for each facet were derived by self-calibrating on selected calibrator sources and then applied to all other sources in the facet. The facet sky models, that were corrected for the direction-dependent effects, were progressively subtracted from the data. The procedure was repeated until only the last facet, containing A1240, was left in the data. The facet calibrator (i.e. B3 1121+435 at RA=11:24:32.043, Dec=+43:15:42.77) that was used to calibrate the facet containing the cluster has a flux of $2.91$ Jy. As the distance between the facet calibrator and A1240 is $14\arcmin$, the ionospheric and instrumental phase corrections should be similar to those towards the direction of A1240.

The data reduction was performed with the facet-calibration pipeline\footnote{\url{https://github.com/tammojan/facet-calibration}}. The pipeline exploits $\mathtt{DPPP}$ (LOFAR Default PreProcessing Pipeline) for data editing (i.e. flagging, averaging, concatenating), $\mathtt{BBS}$  \citep[BlackBoard Seflcal,][]{Pandey2009} for calibration and $\mathtt{WSClean}$ \citep[W-Stacking Clean,][]{Offringa2014} for imaging. To obtain final continuum images of A1240, the calibrated data were deconvolved in $\mathtt{CASA}$ \citep[Common Astronomy Software Applications;][see Subsec. \ref{sec:red_contiuum_imaging}]{Mullin2007}.

\subsection{GMRT 612 MHz}
\label{sec:red_gmrt}

The GMRT $595-629$ MHz observations of A1240 were performed on July 2, 2011 for 6 hours  (project code: 20\_004, PI: A. Bonafede). During the target observations, the radio source 1035+564 was observed for 5 minutes every about $40$ minutes and was used as a phase calibrator. For flux calibration, two bright radio sources 3C 147 and 3C 286 were observed before and after the target observations. The observation details are given in Table \ref{tab:lofar_gmrt_vla_obs}.

The GMRT 612 MHz data were reduced with the Software Peeling and Atmospheric Modelling ($\mathtt{SPAM}$) package \citep{Intema2009a} that performed both direction-independent and direction-dependent calibration. The direction-independent calibration using 3C 147 included flagging RFI, correcting initial phase offsets between the parallel-handed correlations, antenna-based phase delay and amplitude calibration. The flux scale of the primary calibrator 3C 147 was set according to the \cite{Scaife2012} flux scale. Then a phase calibration was performed using a wide-field sky model. In the direction-dependent calibration, $\mathtt{SPAM}$ iteratively solved for amplitude and phase gains towards multiple directions in the target field. The calibration solutions were used to fit a 2D ionospheric model and the derived ionospheric corrections were then applied to the target data. To obtain final images, the direction-dependent calibrated data in the direction of A1240 were deconvolved with $\mathtt{CASA}$ (see Subsec. \ref{sec:red_contiuum_imaging}).

\subsection{VLA 3 GHz}
\label{sec:red_vla}

The VLA S-band observations were performed in C and D configuration on Oct. 10, 2014 and Oct. 09, 2015, respectively  (project: 14B-157, PI: W. Dawson). In each configuration the array was separately pointed at each radio relic. The observations used 16 spectral windows, each of which was split into 64 channels, and covered the $2-4$ GHz bandwidth in total. An overview of the observations is given in Table \ref{tab:lofar_gmrt_vla_obs}.

Following \cite{VanWeeren2016c}, we separately processed the target data for each configuration/pointing with the $\mathtt{CASA}$ package. The target data were Hanning smoothed and corrected for elevation-dependent gains and antenna position offsets. The RFI was flagged with the automatic flaggers in $\mathtt{CASA}$ and $\mathtt{AOFlagger}$ \citep{Offringa2012c}. The antenna delays and bandpass were derived using a model of 3C 286 that is set to the \cite{Perley2013} flux scale and has an uncertainty of a few percent \citep{Perley2013}. The cross-hand delays were solved using 3C 286, assuming a fractional polarization of $11\%$ and a RL-phase difference of $66^\circ$. The polarization leakage terms for every channel were determined from J1407+2827 which served as a low polarization leakage calibrator. After the calibration parameters were derived they were transferred to the target data. The phase calibration of the target field was calculated every 20 minutes using J1146+3958. To improve the fidelity of the target field image, self-calibration loops were then performed. Finally, the calibrated data for the C and D configurations that have the same pointing centres were concatenated in $uv$-plane and were used to make continuum images (see Subsec. \ref{sec:red_contiuum_imaging}).

\subsection{Continuum imaging}
\label{sec:red_contiuum_imaging}

To map the diffuse source structure with the wide-band data sets we exploited multi-scale and multi-frequency synthesis ($\mathtt{MS-MFS}$) in $\mathtt{CASA}$ \citep{Mullin2007,Cornwell2008,Rau2011}. The LOFAR, GMRT and VLA calibrated data were separately $\mathtt{CLEAN}$ed with $\mathtt{MS-MFS}$ to model the complex emission from A1240. The scales used in the deconvolution are  $\mathtt{multiscale}=[0,3,7,25,60,150]\times\text{pixels}$, where the zero scale is for modelling point sources and the larger scales are for sampling the diffuse emission. Due to the wide fractional bandwidth of the VLA observations the primary beam considerably varies across the band and three Taylor terms ($\mathtt{nterms}=3$) were used to model the frequency dependence of the radio emission. $\mathtt{nterms}=2$ and $1$ were used for the LOFAR and GMRT data that have bandwidths of 43 MHz and 33 MHz, respectively. Additionally a wide-field algorithm \citep[$\mathtt{W-projection}$,][]{Cornwell2005,Cornwell2008} was employed to account for the non-coplanarity of the baselines across the sky. Specially depending on the image size, $\mathtt{wprojplanes}=448$ was used for the LOFAR data and $\mathtt{wprojplanes}=384$ was set for the WSRT and GMRT data.

\begin{table*}
	\caption{Imaging parameters that were used  to make images of A1240 and the image properties.}
	\begin{tabular}{lccccccc}
		\hline\hline
		Data          &   $uv$-range   &   $\mathtt{Robust}^a$   &   $\theta_\text{\tiny FWHM}$   &     $\sigma_{\text{\tiny rms}}$      & Stokes &         Fig.         &  \\
		& (k$\lambda$) & ($\mathtt{outertaper}$) & ($\arcsec\times\arcsec$, $PA$) & ($\mu\text{Jy}\,\text{beam}^{-1}$) &        &                      &  \\ \hline
		LOFAR 143 MHz &   $0.2-66$   &  $-0.25$ ($5\arcsec$)   &   $15\times10$ ($87^\circ$)    &                165                 &  $I$   & \ref{fig:hres_lofar} &  \\
		&   $0.2-66$   &  $0.10$ ($25\arcsec$)   &   $41\times36$ ($13^\circ$)    &                410                 &  $I$   &    \ref{fig:lres}    &  \\
		&   $0.2-41$   &  $-0.25$ ($10\arcsec$)  &         $21\times21^b$         &                280                 &  $I$   &  \ref{fig:spx}$^c$   &  \\ \hline
		GMRT 612 MHz  &   $0.2-41$   &  $-0.25$ ($17\arcsec$)  &         $21\times21^b$         &                175                 &  $I$   &  \ref{fig:spx}$^c$   &  \\ \hline
		VLA 3 GHz     &   $0.2-41$   &  $-0.25$ ($9\arcsec$)   &         $21\times21^b$         &                 17                 &  $I$   &  \ref{fig:spx}$^c$   &  \\
		&   $0.2-41$   &         $0.00$          & $18.5\times14.5$ ($85^\circ$)  &                 13                 &  $I$   &    \ref{fig:pols_F}    &  \\
		&   $0.2-41$   &         $0.00$          & $18.5\times14.5$ ($85^\circ$)  &                 $\sim6.4$                 &  $Q$   &  \ref{fig:pols_F}$^d$  &  \\
		&   $0.2-41$   &         $0.00$          & $18.5\times14.5$ ($85^\circ$)  &                 $\sim6.5$                 &  $U$   &  \ref{fig:pols_F}$^d$  &  \\ \hline\hline
	\end{tabular}\\	
	Notes: $^a$: Briggs weighting of visibilities; $^b$: smoothed; $^c$: spectral index map; $^d$: $F$ vector map
	\label{tab:image_para}
\end{table*}

To optimise for imaging on various different angular scales, the LOFAR, GMRT and VLA $uv$-data were weighted using Briggs's $\mathtt{robust}$ weighting \citep{briggs1995} in combination with $\mathtt{uvtapers}$ to down-weighting the outer baselines (see Table \ref{tab:image_para}). The final LOFAR and VLA images were corrected for the attenuation of the primary beams that were generated with $\mathtt{AWImager}$ \citep{Tasse2013} and $\mathtt{CASA}$ \citep{Mullin2007}, respectively. The GMRT 612 MHz images were also corrected for primary beam attenuation\footnote{\href{http://gmrt.ncra.tifr.res.in/gmrt_hpage/Users/doc/manual/Manual_2013/manual_20Sep2013.pdf}{GMRT User's manual}},

\begin{equation}
	A(x) = 1-\frac{3.486}{10^3}x^2+\frac{47.749}{10^7}x^4-\frac{35.203}{10^{10}}x^6+\frac{10.399}{10^{13}}x^8,
\end{equation}
where $x=f\times\theta$, here $\theta$ is angular distance in arcmin to the pointing centre, and $f=0.612$ GHz is the frequency of the GMRT observations.

\subsection{Spectral index measurements}
\label{sec:red_spx_measurements}

To make spectral index maps of A1240 we combined LOFAR 143 MHz, GMRT 612 MHz and VLA 3 GHz continuum images. To measure approximately the same spatial scales of emission, we selected a common $uv$-range ($0.2 - 41.0$ k$\lambda$) for the data sets when making the total intensity images. A common Briggs weighting ($\mathtt{robust}=-0.25$) was applied to the data sets. It is noted that uniform weighting, or attempting to directly match the $uv$-coverage, helps to accurately compare interferometric images. However, such weighting of the uv-data significantly increases the noise levels of the continuum images. Instead, we used the Briggs weighting to increase signal to noise ratio (SNR) of the sources and attempted to ensure that the native resolution of the images from the different arrays was equal by applying different $uv$-tapers. To obtain an angular resolution of $20\arcsec$ we used an outer $uv$-taper of $10\arcsec$, $17\arcsec$ and $9\arcsec$ for the LOFAR, GMRT and VLA data, respectively. The native resolution with these imaging parameters was $19.7\arcsec\times14.9\arcsec$ (position angle of $PA=86.3^{\circ}$) for the LOFAR image,  $19.2\arcsec\times14.0\arcsec$ ($PA=24.4^{\circ}$) for the GMRT image and $20.0\arcsec\times15.8\arcsec$ ($PA=-84.4^{\circ}$) for the VLA image. These total intensity maps were smoothed to an identical resolution of $21\arcsec$, aligned and regrided. The LOFAR, GMRT and VLA images have noise levels of $\sigma_{\text{\tiny rms}}=280$, 175 and 17 $\mu$Jy/beam, respectively. The spectral indices were calculated for each pixel by fitting the $\geqslant3\sigma_{\text{\tiny rms}}$ pixels in at least two images with a power-law function, $S\propto\nu^\alpha$. To estimate the spectral index error, we adapted a common flux scale uncertainty of $10\%$ associated with the calibration of the LOFAR, GMRT and VLA data, as commonly used in the literature \citep[e.g.][]{VanWeeren2016b,Hoang2017a}.

\subsection{Polarization measurements}
\label{sec:red_pol_measurements}

We used the VLA $2-4$ GHz data to study the linear polarization properties of the faint diffuse emission from A1240. We made multiple polarization maps with (i) the full bandwidth $2-4$ GHz data which maximised the polarization detection significance, (ii) successive narrower (480 MHz) band data to examine the frequency dependence of the polarized emission and (iii) successive 224 MHz bandwidth chunks to ensure that our measurements were not suffering from bandwidth depolarisation. In each case we made Stokes $I$, $Q$ and $U$ images with $\mathtt{WSClean}$ \citep{Offringa2014}. The imaging was done with the multi-scale and joined-channel deconvolution algorithm \citep{Offringa2017}. We also used Briggs ($\mathtt{robust}=0.00$) weighting on the $uv$-data. The reason for using $\mathtt{WSClean}$, instead of $\mathtt{CASA}$, is because the combination of multiscale and Stokes $Q$/$U$ $\mathtt{CLEAN}$, which is essential for recovering the faint diffuse polarized emission of A1240, is not yet available in $\mathtt{CASA}$ (version 4.7). To obtain the polarization intensity $P$ and angle $\phi$ maps, the Stokes $Q$ and $U$ images were combined as follows,

\begin{equation}
	P = \sqrt{Q^2+U^2}; \quad  \phi = \frac{1}{2}\arctan\frac{U}{Q}.
	\label{eq:Pols_cal}
\end{equation}

From the polarized $P$ and Stokes $I$ emission maps, the total polarization fraction, $F=P/I$, was calculated for pixels within the $\geqslant3\sigma_{\text{\tiny rms}}$ region of the Stokes $I$ image. To obtain the corrected flux measurements, the final Stokes $P$ and $I$ images were then divided by the VLA primary beam to correct for the sensitivity attenuation. 

The polarization angle calculated from Eq. \ref{eq:Pols_cal} was further corrected for the Faraday rotation caused by the Galactic magnetic field (i.e. $\phi_{\text{\tiny A1240}}=\phi_{\text{\tiny Eq.\,\ref{eq:Pols_cal}}}-\phi_{\text{\tiny Galactic}}$). Given the mean Galactic rotation measure ($\mathtt{RM}$) of $9.4\,\text{rad/m}^2$ towards the direction of A1240 \citep{Oppermann2012}, the Galactic Faraday rotation ($\phi_{\text{\tiny Galactic}}=\mathtt{RM}\times\lambda^2$) is $12^\circ$ and $3^\circ$ at the lower and higher edges of the $2-4$ GHz band, respectively. Since the polarized emission map was made with full bandwidth data that has the central frequency at 3 GHz, we corrected the Galactic Faraday rotation using $\phi_{\text{\tiny Galactic}}=5^\circ$  \citep{Oppermann2012}.

\subsection{\textit{Chandra} X-ray data}
\label{sec:red_chandra}

The \textit{Chandra} ACIS-I observation (ID: \#4961, PI: Kempner) of A1240 was taken on Feb. 5, 2005 and has a duration of 52 ks. Following the data reduction procedure described in \cite{Vikhlinin2005}, we applied the calibration files\footnote{CIAO v4.6 and CALDB v4.6.5} using the $\mathtt{chav}$ package\footnote{\url{http://hea-www.harvard.edu/\~alexey/CHAV}}. The calibration includes filtering out bad pixels, correcting for the position-dependent inefficiency of the charge transfer and correcting for photon energies with gain maps. The background emission was subtracted using standard blank sky files. For more details on the reduction procedure, we refer to \cite{Vikhlinin2005}.

\section{Results}
\label{sec:results}

\begin{figure*}
	\includegraphics[width=0.6\textwidth]{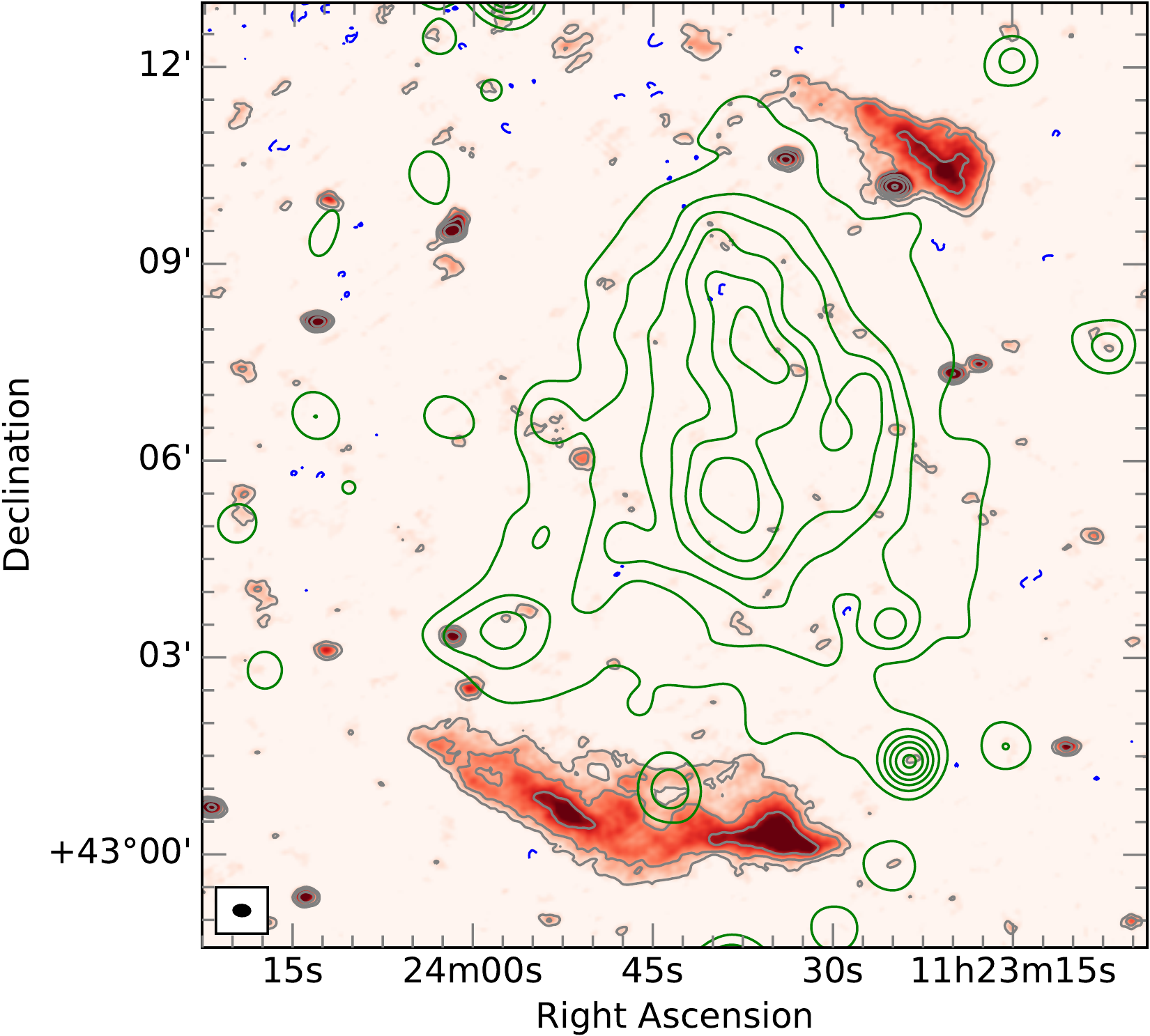}
	
	\caption{LOFAR 143 MHz total intensity map of A1240 with contours in grey (positive) and blue (negative) ($\theta_{\text{\tiny FWHM}}=15\arcsec\times10\arcsec$, $PA=87^\circ$). The contours are  $[-3,3,6,12,24,48]\times\sigma_{\text{\tiny rms}}$, where $\sigma_{\text{\tiny rms}}=165\,\text{\ensuremath{\mu}Jy/beam}$. The green contours are the \textit{Chandra} X-ray surface brightness smoothed with a 2D Gaussian kernel to $40\arcsec$ resolution. The X-ray contour levels are $[3,6,9,12,15,18]\times\sigma$, where $\sigma=0.5\times 10^{-6}\,\text{cts/s/arcmin}^2$.}
	
	\label{fig:hres_lofar}
\end{figure*}

\begin{figure}
	\includegraphics[width=1\columnwidth]{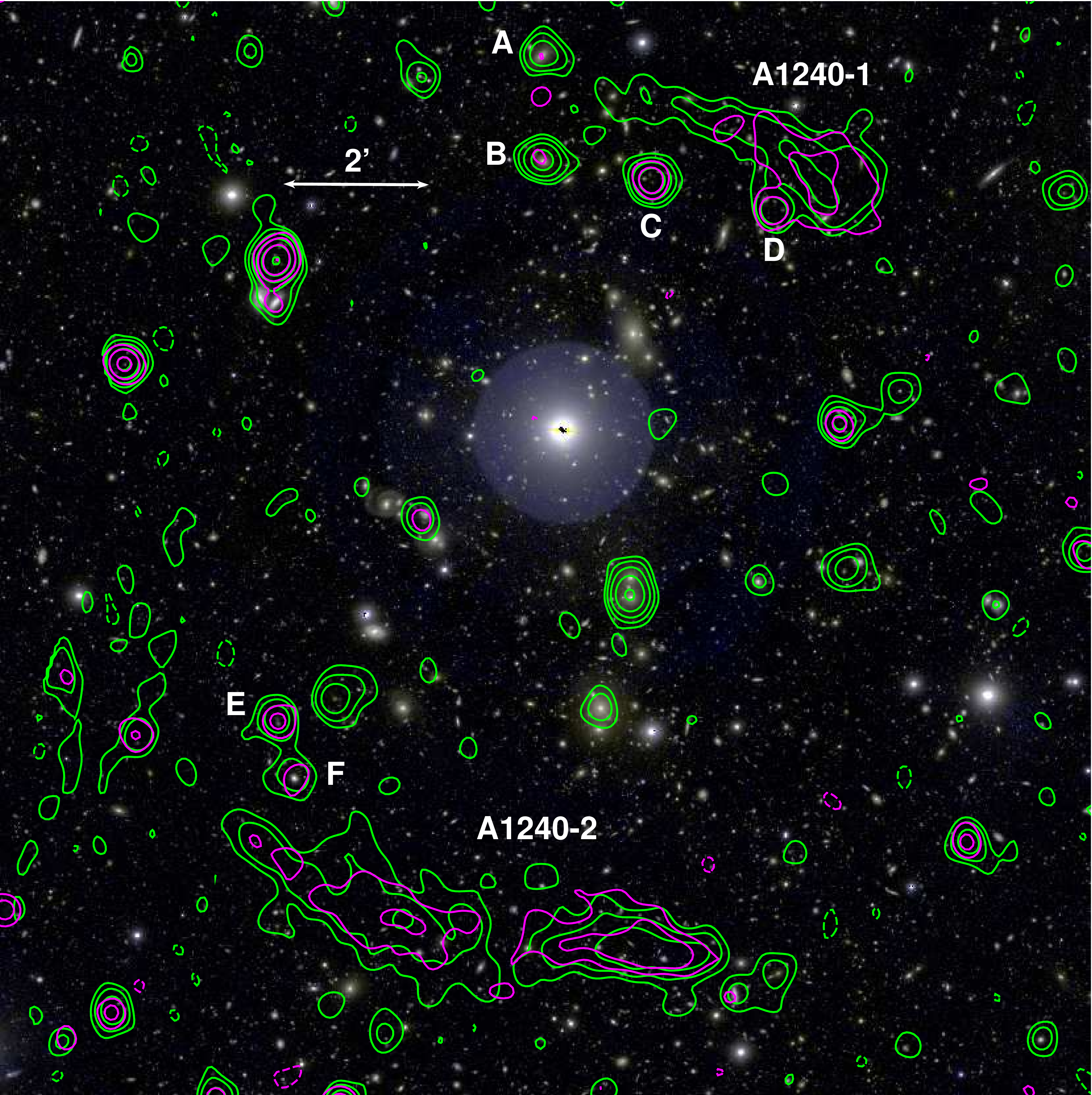}
	
	\caption{A Subaru r,g band image of A1240. The VLA (green) and GMRT (magenta) contours are levelled at $[-3,3,6,12,24,48]\times\sigma_{\text{\tiny rms}}$ (dashed negative), where $\sigma_{\text{\tiny rms}}=175\,\text{\ensuremath{\mu}Jy/beam}$ and $17\,\text{\ensuremath{\mu}Jy/beam}$ for the GMRT and VLA images, respectively. The resolution of the radio images is $\theta_{\text{\tiny FWHM}}=21\arcsec\times21\arcsec$. The radio sources are labelled.}
	\label{fig:hres_vla_gmrt}
\end{figure}

\begin{figure*}
	\includegraphics[width=0.6\textwidth]{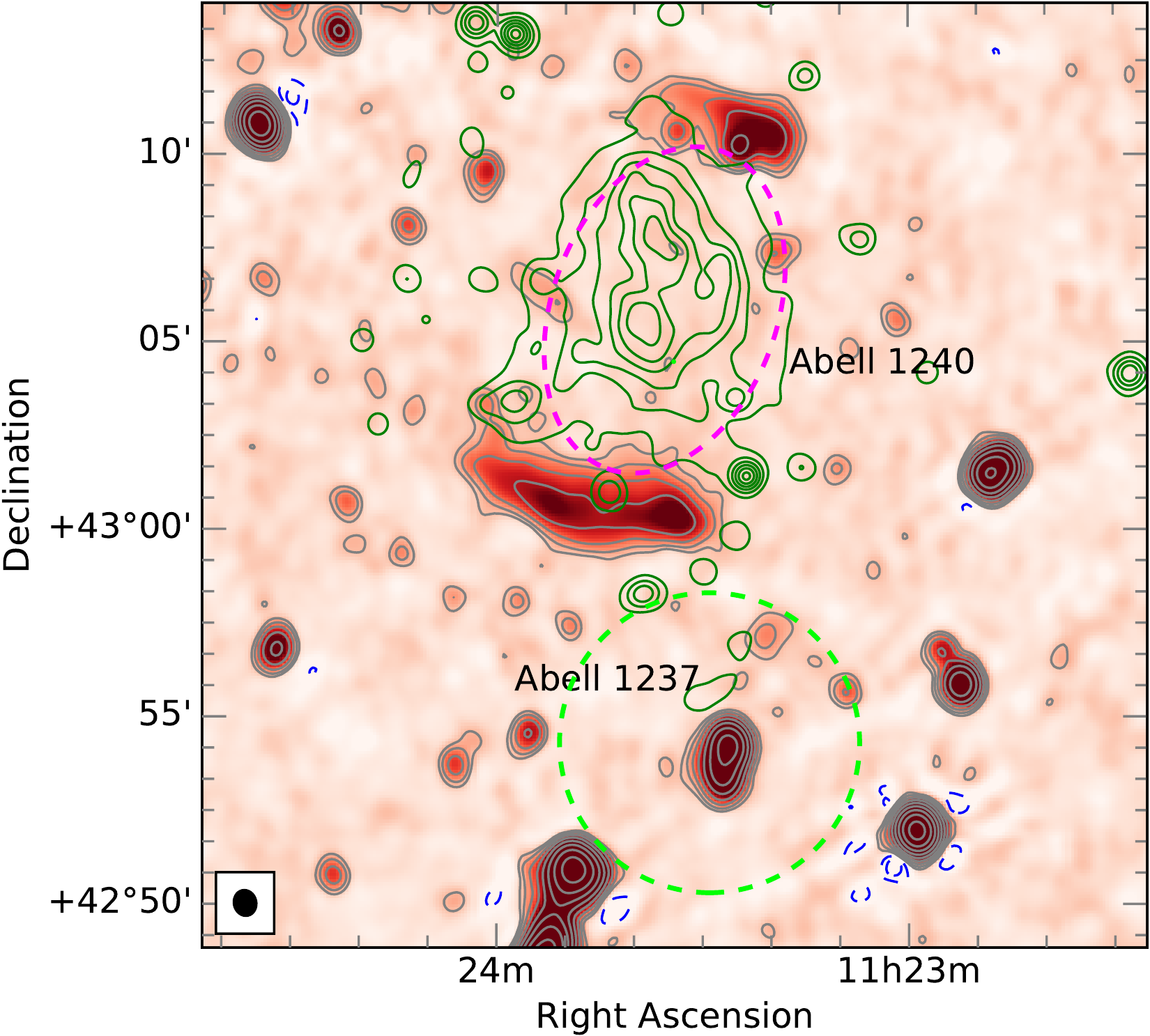}
	
	\caption{LOFAR 143 MHz total intensity map of A1240 with contours in grey (positive) and blue (negative) ($\theta_{\text{\tiny FWHM}}=41\arcsec\times36\arcsec$, $PA=13^\circ$). The contours are  $[-3,3,6,12,24,48]\times\sigma_{\text{\tiny rms}}$, where $\sigma_{\text{\tiny rms}}=410\,\text{\ensuremath{\mu}Jy/beam}$. The X-ray contours are identical to those in Fig. \ref{fig:hres_lofar} and are only available for A1240. The dashed magenta ellipse shows the region where the upper limit of diffuse emission is estimated in Subsec. \ref{sec:dics_radio_halo}.  The green dashed circle marks the region of A1237.}
	
	\label{fig:lres}
\end{figure*}

In Fig. \ref{fig:hres_lofar} we present high-resolution continuum images of A1240 that were made with LOFAR at a frequency of 143 MHz. The GMRT 612 MHz and VLA 3 GHz contours are overlaid on the Subaru r-band image in Fig. \ref{fig:hres_vla_gmrt}. The resolution is $\theta_{\text{\tiny FWHM}}=15\arcsec\times10\arcsec$ ($PA=87^\circ$) for the LOFAR image and is  $\theta_{\text{\tiny FWHM}}=21\arcsec\times21\arcsec$ for the VLA and GMRT images. A common Briggs' $\mathtt{robust}$ weighting of $-0.25$ and $\mathtt{outertaper}=5\arcsec$, $17\arcsec$ and $9\arcsec$ were used for the LOFAR, GMRT and VLA imaging (see Subsec. \ref{sec:red_contiuum_imaging}). In Fig. \ref{fig:lres} we present a low-resolution ($41\arcsec\times36\arcsec$, $PA=13^{\circ}$) 143 MHz image of A1240 ($\mathtt{robust}=0.10$, $\mathtt{outertaper}=25\arcsec$). The radio relics in the northern and southern outskirts of A1240, that were previously observed with the WENSS 325 MHz in  \cite{Kempner2001} and with the VLA 325 MHz and 1.4 GHz in \cite{Bonafede2009}, were detected with a peak flux of $12\sigma_{\text{\tiny rms}}$ in our high resolution images (Figs. \ref{fig:hres_lofar}-\ref{fig:lres}). \cite{Bonafede2009} previously presented spectral index and polarization properties and equipartition magnetic field measurements of the relics. With the wide-band observations between 143 MHz and 3 GHz we provide new measurements on the spectra and polarization properties of the relics.

\subsection{The double radio relics}
\label{sec:res_radio_relics}

The detected relics (Figs. \ref{fig:hres_lofar} and \ref{fig:hres_vla_gmrt}) have projected sizes of $0.68\times0.20\,\text{Mpc}^2$ and $1.35\times0.35\,\text{Mpc}^2$, respectively; and their major axes are aligned with the $3\sigma_\text{\tiny rms}$ edges of the \textit{Chandra} X-ray emission (Fig. \ref{fig:lres}). For both relics, the surface brightness gradient is steeper on the outer edges than that in the inner edges. Across the length of A1240-1, the surface brightness gradually decreases from west to east. The emission on the western and eastern sides of A1240-2 appears completely detached at 612 MHz and 3 GHz (Fig. \ref{fig:hres_vla_gmrt}), but is connected at 143 MHz (Fig. \ref{fig:hres_lofar}).

\subsubsection{Spectral analysis}
\label{sec:res_spx_analysis}

\begin{figure*}
	\centering
	\noindent\begin{minipage}[t]{0.34\textwidth}%
		\begin{center}
			\includegraphics[clip,width=1\textwidth]{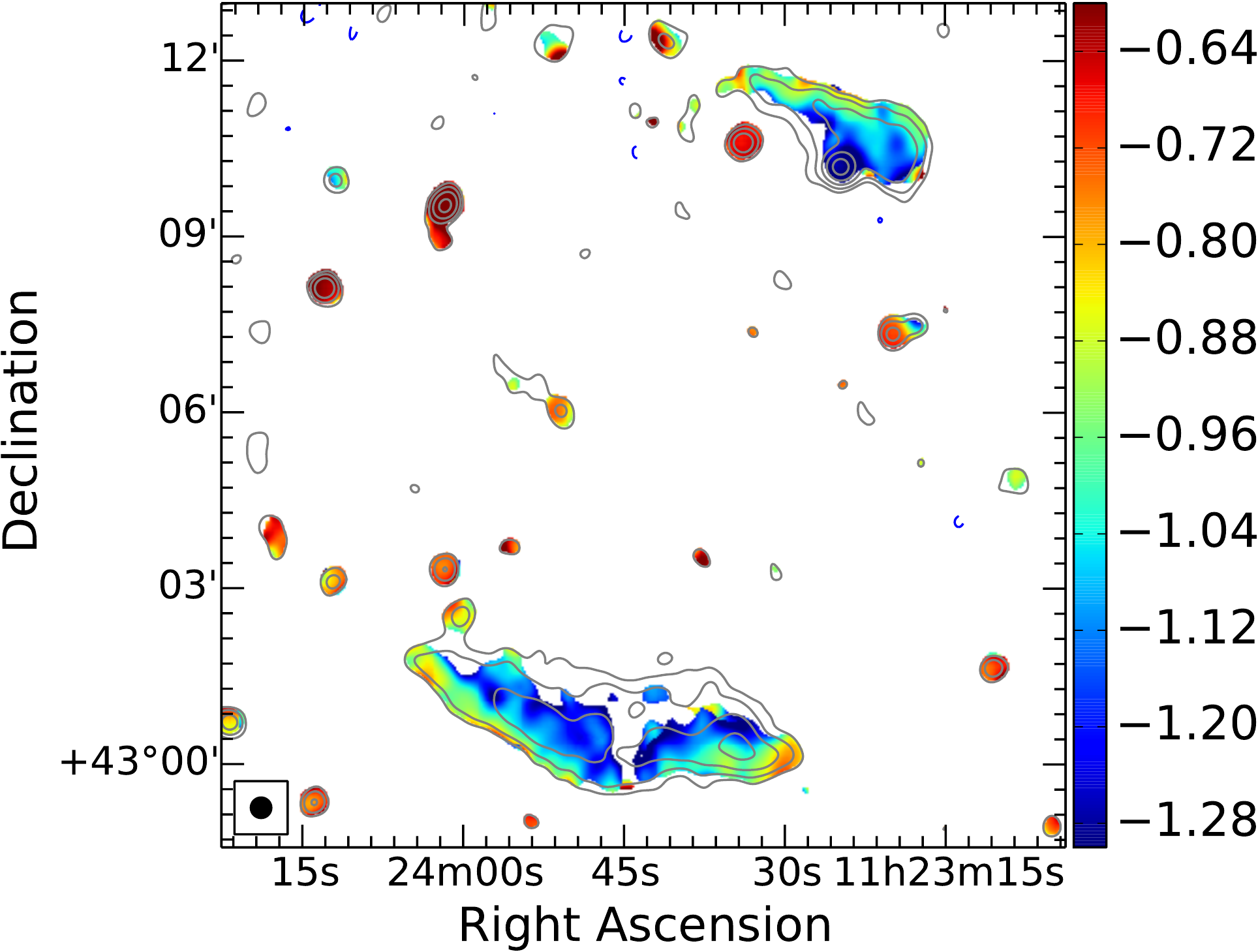}
			\par\end{center}%
	\end{minipage}%
	\hfill
	\noindent\begin{minipage}[t]{0.33\textwidth}%
	\begin{center}
		\includegraphics[clip,width=1\textwidth]{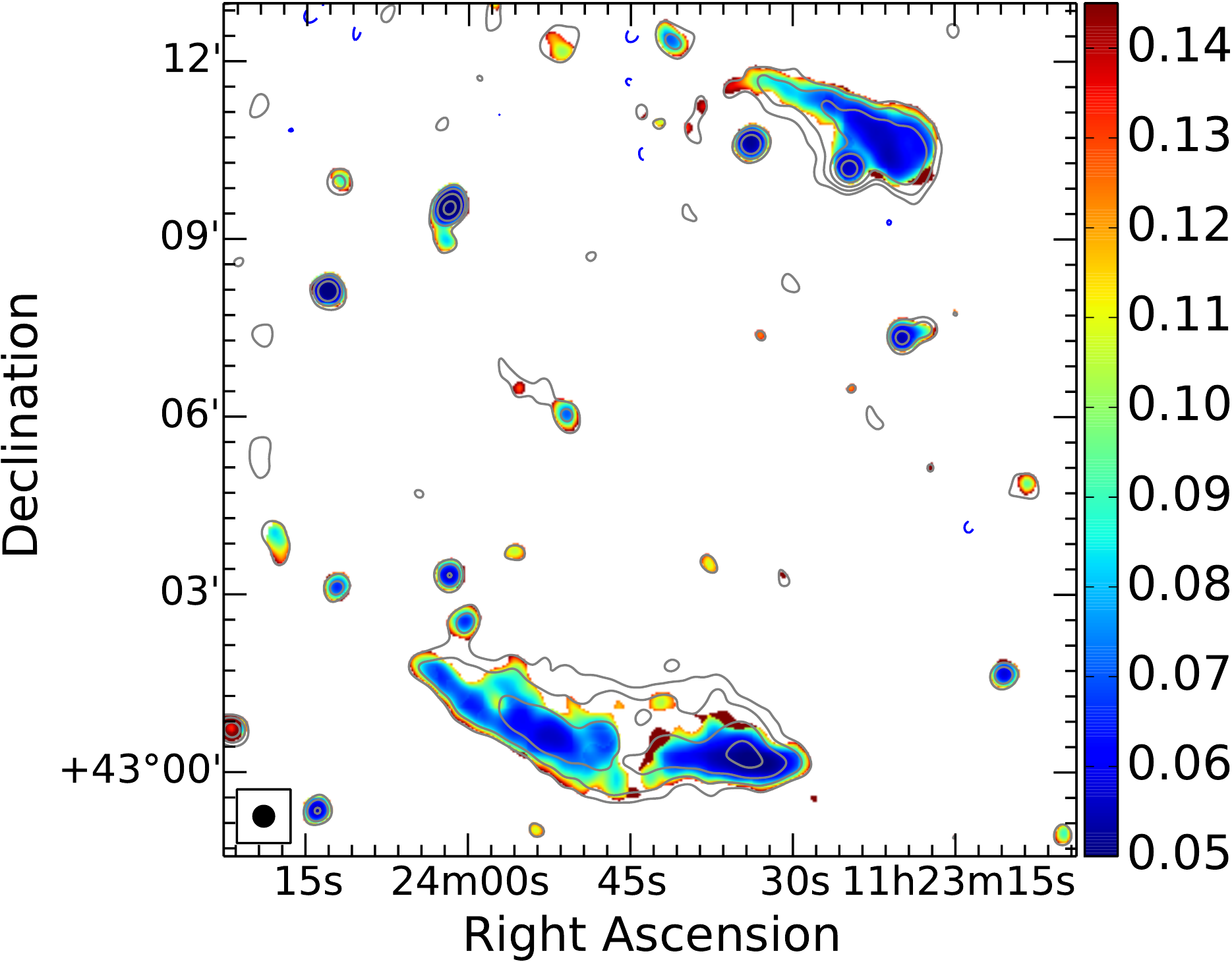}
		\par\end{center}%
	\end{minipage}%
	\hfill	
	\noindent\begin{minipage}[t]{0.28\textwidth}
		\centering
		\begin{tikzpicture}
		\node[anchor=south west,inner sep=0] (image) at (0,0) {\includegraphics[width=1\columnwidth]{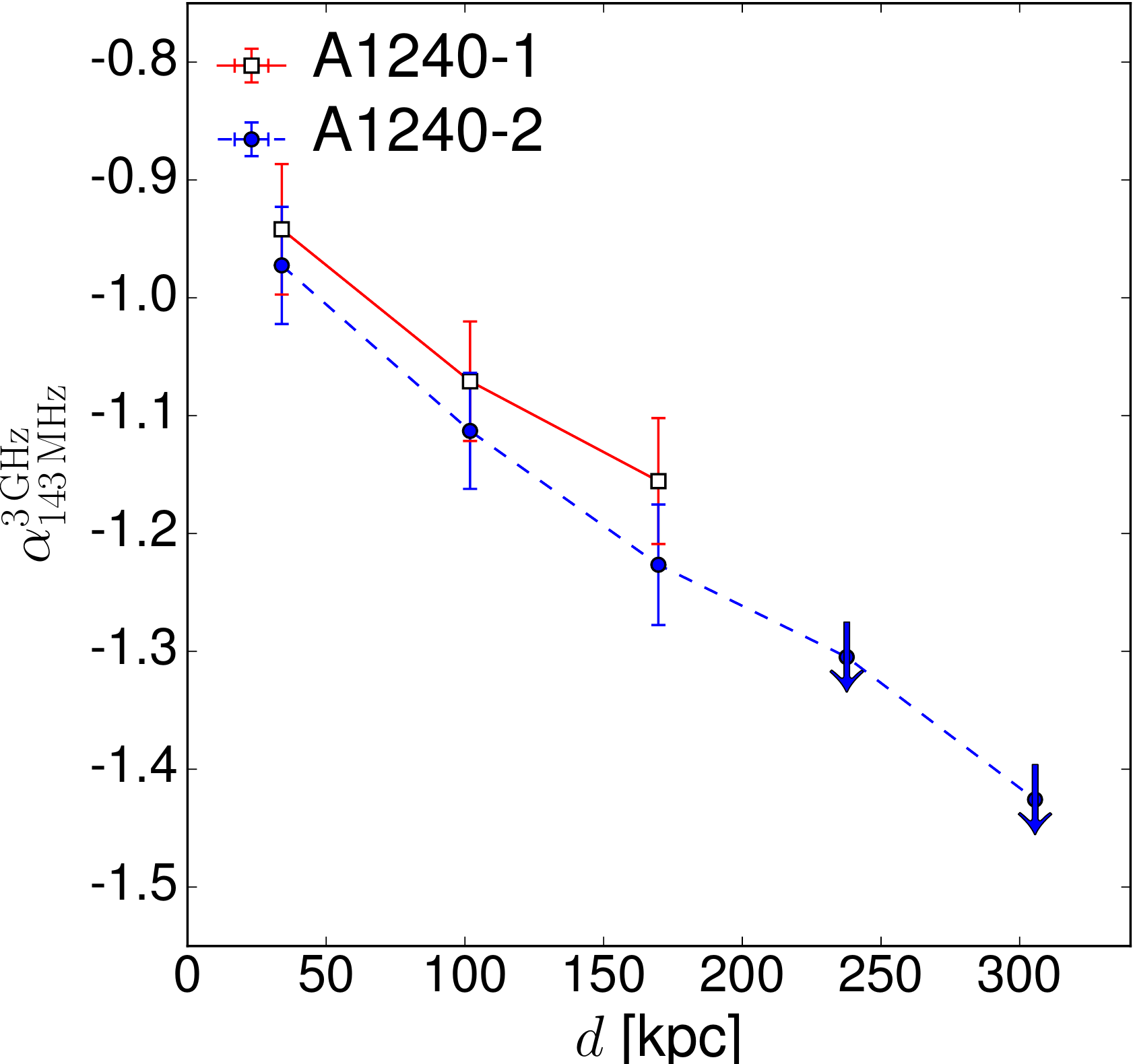}};
		\begin{scope}[x={(image.south east)},y={(image.north west)}]
		
		\node[anchor=north east,inner sep=0] (image) at (0.985,0.985) {\includegraphics[width=0.4\columnwidth]{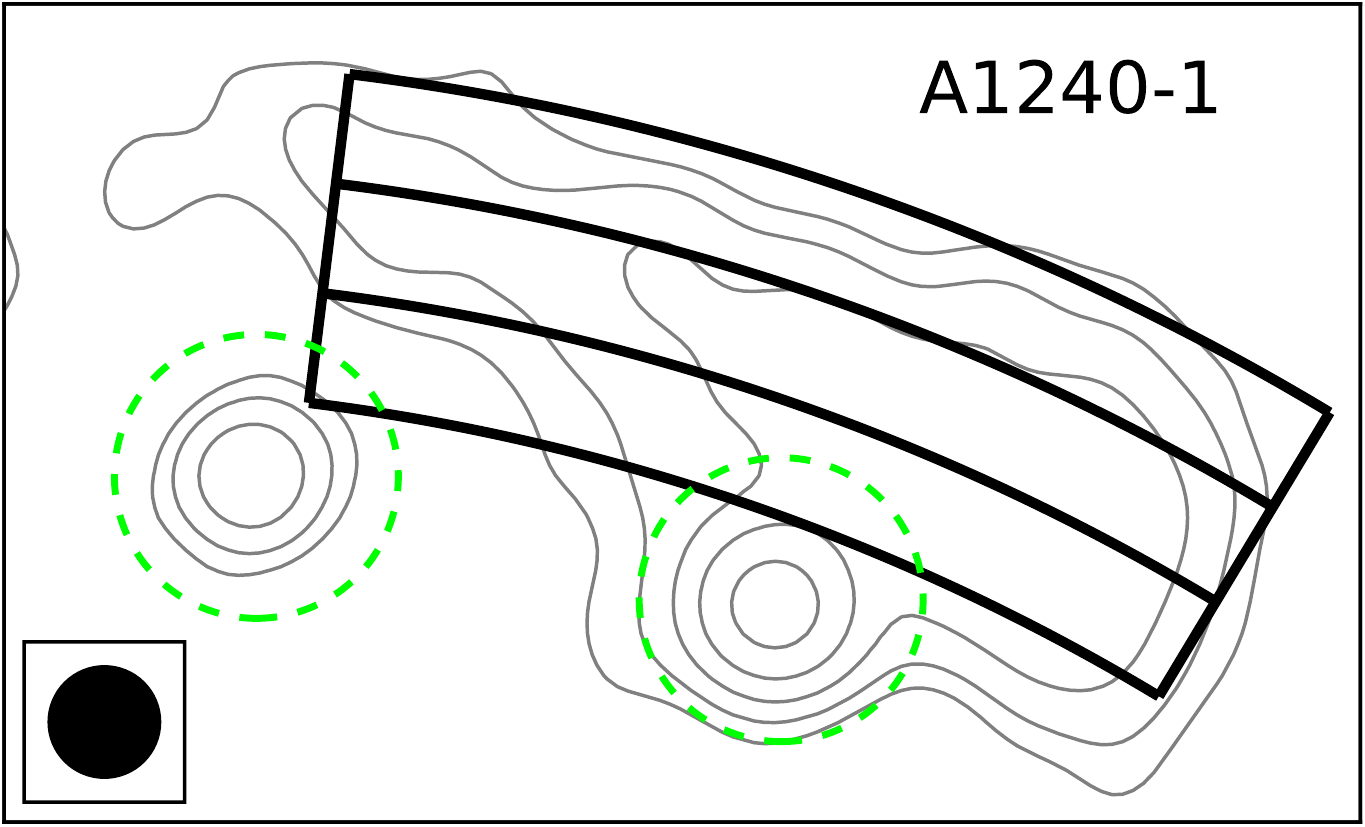}};
		
		\node[anchor=south west,inner sep=0] (image) at (0.18,0.12) {\includegraphics[width=0.45\columnwidth]{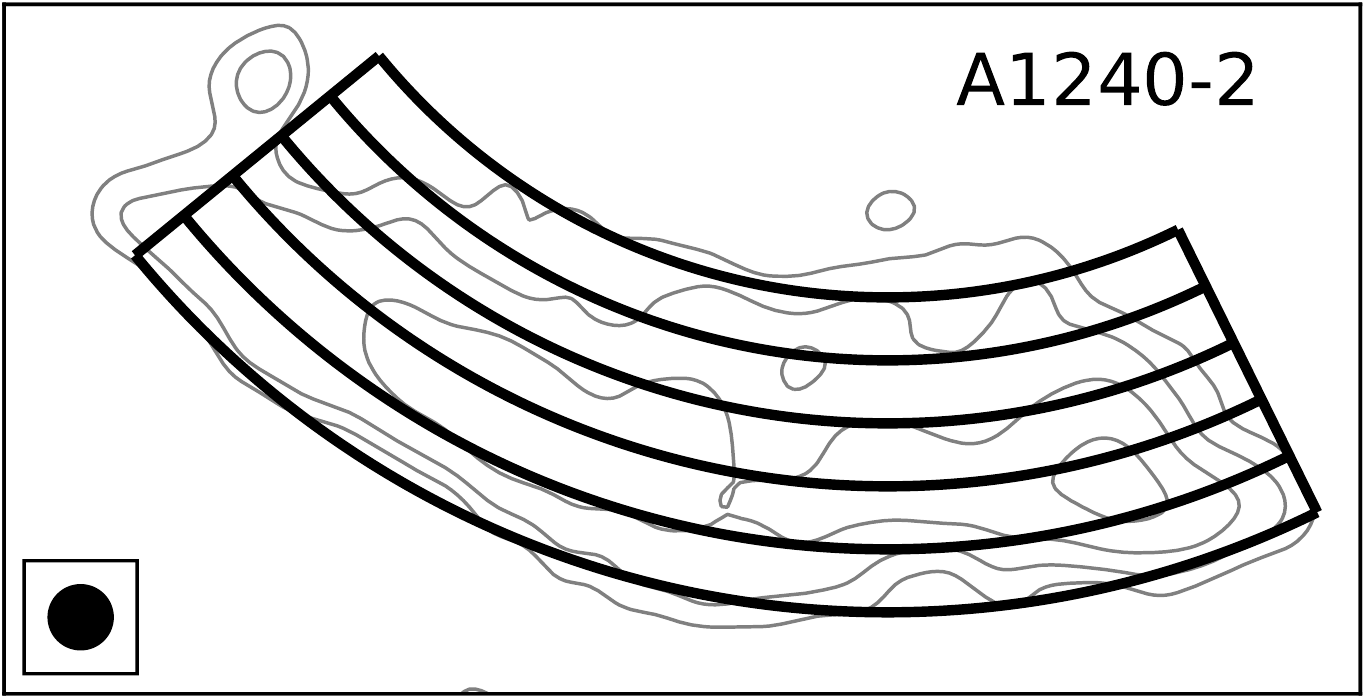}};
		
		\end{scope}
		\end{tikzpicture}
	\end{minipage}\caption{Left: Three-frequency spectral index map between 143 MHz and 3 GHz of A1240 at $21\arcsec$ (or $\sim68$ kpc) resolution. Middle: The corresponding spectral index error map. Right: The spectral index profiles across the width of the relics A1240-1 and A1240-2 and towards the cluster centre. The flattest spectral indices are $-0.94\pm0.06$ and $-0.97\pm0.05$ at the outer edges of A1240-1 and A1240-2, respectively. The subplots show the regions where the spectral indices were extracted. The compact sources (i.e. green dotted circles) were masked. The radial size of the region is equal to the synthesized beam size of $21\arcsec$. The downward pointing arrows indicate the upper limit of the spectral indices that have $<2\sigma_{\text{\tiny rms}}$ detection confidence levels in VLA and/or GMRT observations. The LOFAR 143 MHz superimposed contours in both panels are at identical spacings to those in Fig. \ref{fig:hres_lofar} (here $\sigma_{\text{\tiny rms}}=280\,\text{\ensuremath{\mu}Jy/beam}$).
	\label{fig:spx} }
\end{figure*}
 
In Fig. \ref{fig:spx} (left) we present the three-frequency spectral index map between 143 MHz and 3 GHz of A1240 (see Subsec. \ref{sec:red_spx_measurements}). In Fig. \ref{fig:spx} (left), the spectral indices for A1240-1 and A1240-2 were found to steepen from the outer edge towards the inner regions. The steepening trend across the width of the relics is better visible in the spectral index profiles in the right panel of Fig. \ref{fig:spx}. In particular the spectral indices are $-0.94\pm0.06$ and $-0.97\pm0.05$ at the outer edges of A1240-1 and A1240-2, respectively.  Towards the inner regions at a distance of $63\arcsec$ ($\sim204$ kpc) from the outer edges of A1240-1 and A1240-2, the spectral indices significantly steepen to $-1.16\pm0.05$ and $-1.23\pm0.05$, respectively.

\begin{figure}
	\includegraphics[width=1\columnwidth]{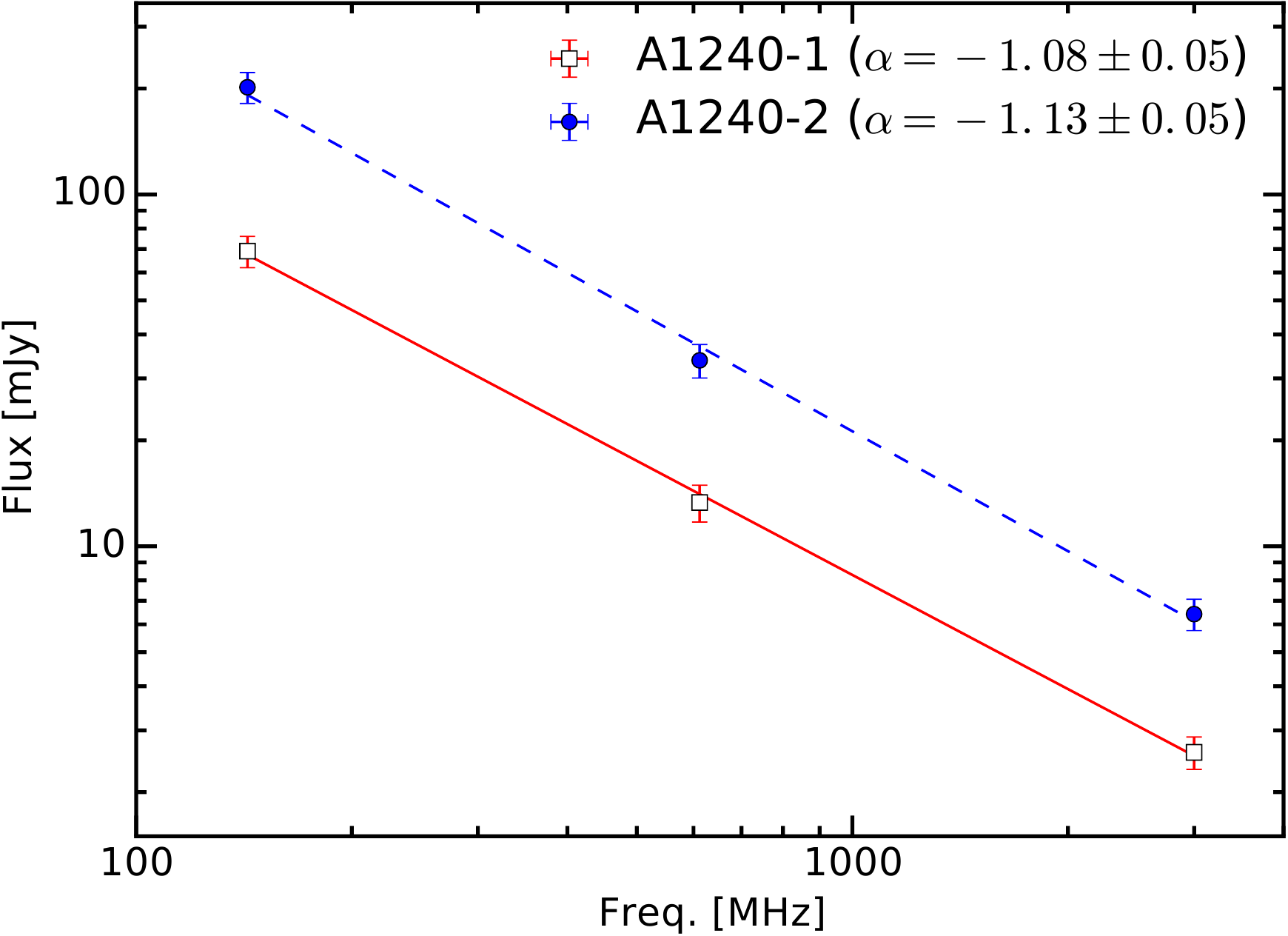}
	
	\caption{Integrated spectra for the radio relics of A1240. The integrated fluxes of the relics were measured in the LOFAR 143 MHz, GMRT 612 MHz and VLA 3 GHz $21\arcsec$-resolution images (Table \ref{tab:image_para}) and are given in Table \ref{tab:int_fluxes}. The spectral index values that were obtained from the spectral power-law fitting, $S\propto\nu^\alpha$, for the relics are given in Table \ref{tab:Int_Spx_Mach}.}
	
	\label{fig:int_spx}
\end{figure}

To estimate the integrated spectral indices of A1240-1 and A1240-2, we used the LOFAR 143 MHz, GMRT 612 MHz and VLA 3 GHz images that were used to make the spectral index map in Fig. \ref{fig:spx}. The integrated fluxes were measured within the $\geqslant3\sigma_{\text{\tiny rms}}$ region of the LOFAR image on all three images, are given in Table \ref{tab:int_fluxes} and is plotted in Fig. \ref{fig:int_spx}. The measured fluxes at three frequencies were fit with a power-law function, $S\propto\nu^\alpha$. The best-fit spectral indices for A1240-1 and A1240-2 were estimated to be $-1.08\pm0.05$ and $-1.13\pm0.05$, respectively. Our spectral index measurements are statistically consistent with the 325 MHz $-$ 1.4 GHz measurements of $-0.96\pm0.26$ and $-1.11\pm0.27$ in \cite{Kempner2001} and $-1.2\pm0.1$ and $-1.3\pm0.2$ in \cite{Bonafede2009}. In addition, to search for spectral curvature we divided the data into two frequency intervals: from 143 to 612 MHz and from 612 MHz to 3 GHz. We estimated the integrated spectral indices between 143 and 612 MHz to be $-1.13\pm0.11$ and $-1.23\pm0.10$ for A1240-1 and A1240-2, and between 612 MHz and 3 GHz to be $-1.03\pm0.10$ and $-1.08\pm0.05$ for A1240-1 and A1240-2, providing no clear evidence for spectral curvature between 143 MHz and 3 GHz in either relic.

\begin{table}
	\centering{}
	\caption{Integrated fluxes for the radio relics of A1240.}
	\begin{tabular}{lccc}
		\hline\hline
		Source  & $S_\text{\tiny 143 MHz}$ (mJy)& $S_\text{\tiny 612 MHz}$ (mJy)& $S_\text{\tiny 3 GHz}$ (mJy) \\\hline
		A1240-1 &      $68.45\pm1.38$      &      $13.32\pm1.60$      &     $2.54\pm0.09$      \\
		A1240-2 &     $202.39\pm2.40$      &      $33.77\pm3.70$      &     $6.43\pm0.17$      \\ \hline\hline
	\end{tabular} \\
	\label{tab:int_fluxes}
\end{table}

\subsubsection{Polarization analysis}
\label{sec:res_pol_analysis}

\begin{figure*}
	\centering
	\begin{minipage}{0.48\textwidth}%
		\includegraphics[width=1\columnwidth]{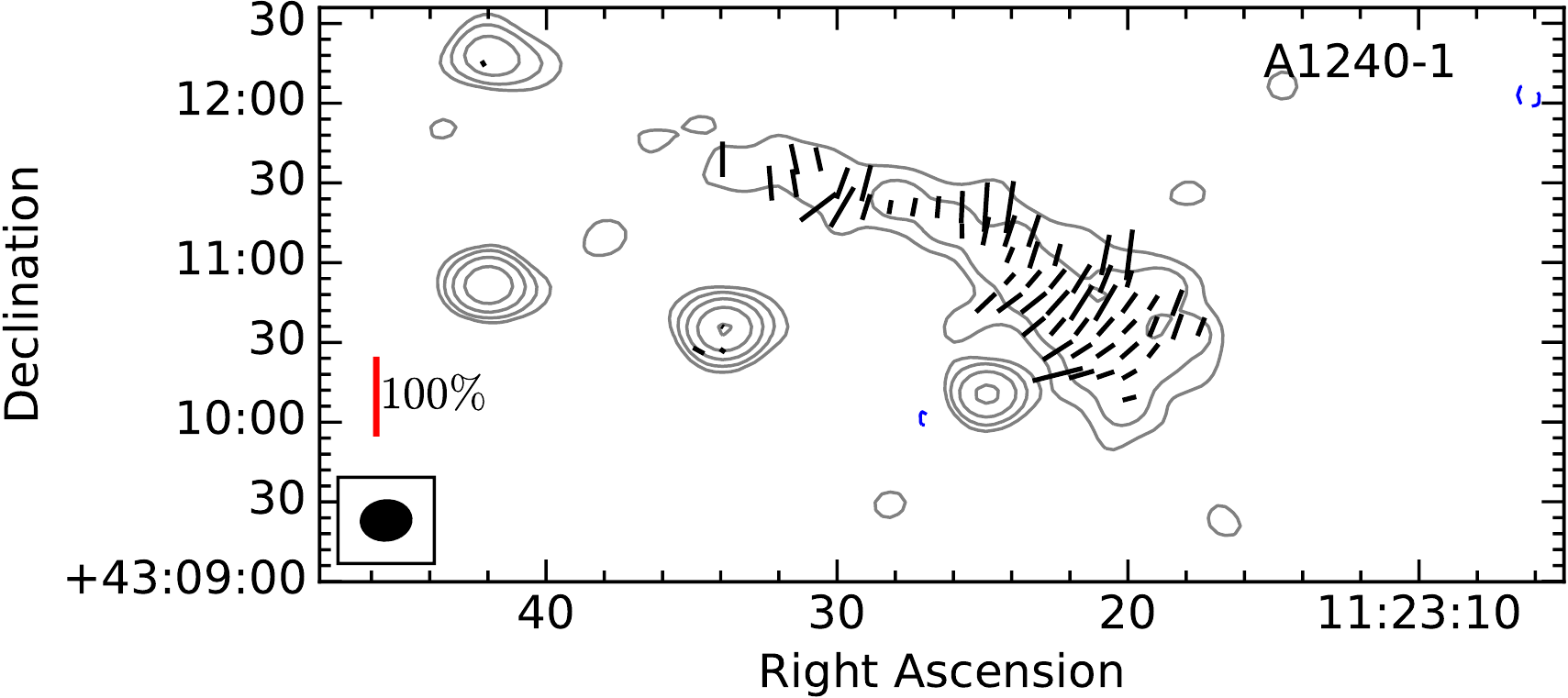} \\		
		\includegraphics[width=1\columnwidth]{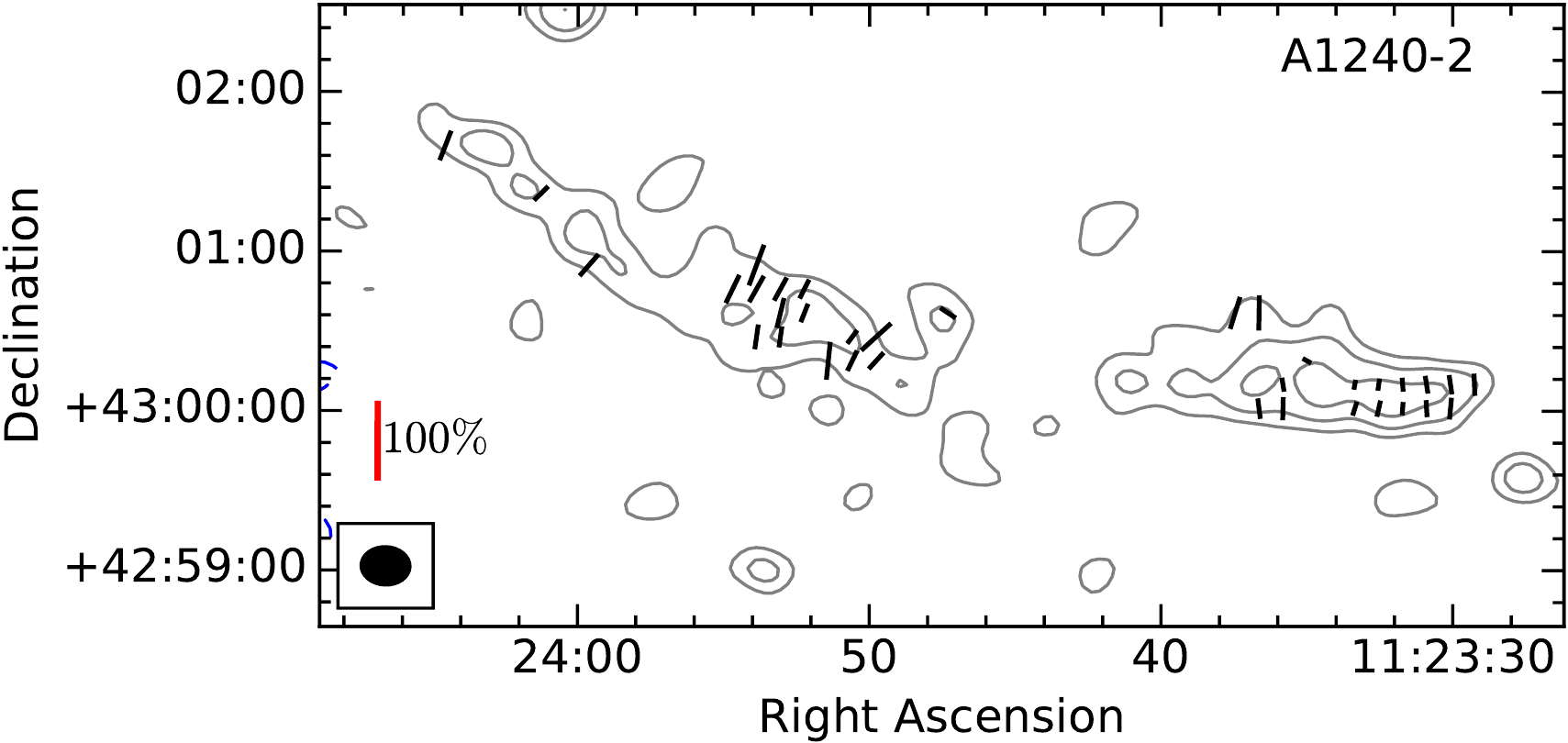}%
	\end{minipage}\hfill{}%
	\begin{minipage}{0.48\textwidth}%
		\includegraphics[width=1\columnwidth]{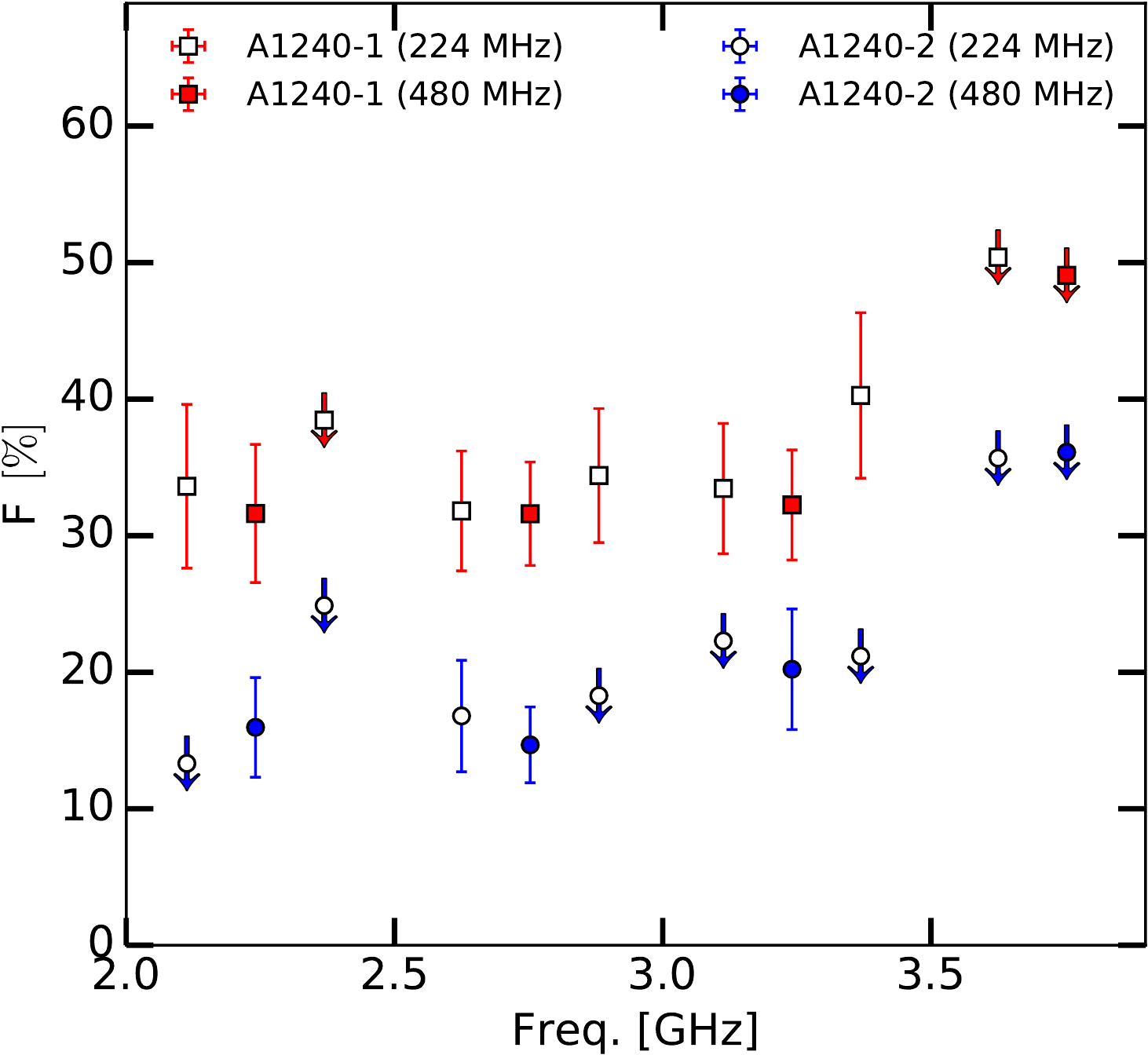}%
		
	\end{minipage}\caption{Left: Electric field vector maps in the regions of the A1240 relics. The red vertical reference lines for 100\% of fractional polarization are shown in the left bottom corners. The VLA $2-4$ GHz $18.5\arcsec\times14.5\arcsec$ resolution (grey) contours are at identical levels to those in Fig. \ref{fig:hres_lofar} (here $\sigma_{\text{\tiny rms}}=13\,\mu\text{Jy/beam}$). Right: Fractional polarization of A1240-1 and A1240-2 between 2 and 4 GHz. The down pointing arrows indicate the data points where polarized emission is below $1.4\,\sigma_\text{\tiny rms}$ detection limit. The fractional polarization for the VLA 224 MHz and 480 MHz bandwidth data sets are in line with each other.}
	\label{fig:pols_F}
\end{figure*}

In Fig. \ref{fig:pols_F} we present polarization electric field vectors in the regions of the relics. The polarized radio emission is extended along the length of the relics. The electric field vectors are approximately perpendicular to the major axes of the relics. The integrated fractional polarization at 3 GHz is $29\pm2\%$ (up to $\bar{F}_\text{\tiny beam}=58\%$ in the most polarised regions in Fig. \ref{fig:pols_F}) for A1240-1 and is $16\pm2\%$ (up to $\bar{F}_\text{\tiny beam}=40\%$) for A1240-2. Our fractional polarization measurement is close to the value of $26\%$ for A1240-1 measured from the VLA 1.4 GHz data \citep{Bonafede2009}, but lower than that of $29\%$ for A1240-2.

We examined the dependence of the fractional polarization on frequency. The VLA $2-4$ GHz data were split into 4 sub data sets, each of which has a bandwidth of 480 MHz. For each 480 MHz data set, we made polarized emission and total intensity (Stokes $I$) images in a similar manner to the procedure used for the full-band $2-4$ GHz data (see Subsec. \ref{sec:red_pol_measurements}). The region that was used to measure the integrated polarized fluxes is within the $\geqslant3\sigma_{\text{\tiny rms}}$ pixels of the full-band Stokes $I$ image. The fractional polarization, $F=P/I$, was estimated for each 480 MHz data set and is plotted in Fig. \ref{fig:pols_F}. The mean polarized emission for the first-three 480 MHz data chunks for A1240-1 and A1240-2 were measured to be $\sim32\pm4\%$ and $\sim17\pm4\%$, respectively. These fractional polarization measurements are consistent with the mean values (i.e. $29\pm2\%$ for A1240-1 and $16\pm2\%$ for A1240-2) that we measured directly from the full-band polarized emission map (Fig. \ref{fig:pols_F}) indicating that our measurements are not severely affected by bandwidth depolarisation. Furthermore, measurements were also made with a bandwidth of 224 MHz and these, whilst at lower SNR, are consistent with both the 2 GHz and 480 MHz bandwidth measurements (Fig. \ref{fig:pols_F}). The polarization angle of the emission over the regions of A1240-1 and A1240-2 was measured to be approximately constant over the $2-4$ GHz bandwidth. 

\begin{table*}
	\centering{}
	\caption{Spectral properties and Mach numbers for the radio relics.}
	\begin{tabular}{lccccccc}
		\hline\hline
		Source  & $\alpha_\text{\tiny int}$ & $\alpha_\text{\tiny inj}$ & $\alpha_\text{\tiny int}^a$ & $\mathcal{M_\text{\tiny int}}$ & $\mathcal{M_\text{\tiny inj}}$ & $\mathcal{M_\text{\tiny int}}^a$ &  \\ \hline
		A1240-1 &      $-1.08\pm0.05$       &      $-0.94\pm0.06$       &        $-1.2\pm0.1$         &      $5.1^{+3.1}_{-1.1} $      &          $2.4\pm0.1$           &           $3.3\pm0.2$            &  \\
		A1240-2 &      $-1.13\pm0.05$       &      $-0.97\pm0.05$       &        $-1.3\pm0.2$         &      $4.0^{+1.1}_{-0.6} $      &          $2.3\pm0.1$           &           $2.8\pm0.3$            &  \\ \hline\hline
	\end{tabular} \\
	Notes: Col. 1: source name; Col. 2: integrated spectral index between 143 MHz and 3 GHz (Subsec. \ref{sec:res_spx_analysis}); Col. 3: injection spectral index calculated in the outer edge regions (Subsec. \ref{sec:res_spx_analysis}); Col. 4: integrated spectral index between 325 MHz and 1.4 GHz \citep{Bonafede2009}; Col. $5-7$: Mach numbers derived from Col. $2-4$, respectively; $^a$: data from \cite{Bonafede2009}.
	\label{tab:Int_Spx_Mach}
\end{table*}

\subsection{A connection with A1237}
\label{sec:res_a1237}

Fig. \ref{fig:lres} shows the location of A1237 which is a cluster that is falling to A1240 from the south \citep{Barrena2009}. In the central region of A1237 we detect a tailed radio galaxy that shows extended emission towards the south, suggesting that the radio galaxy is moving to the north with respect to the local ICM. No diffuse large-scale emission was observed from the ICM of A1237 or the region between the clusters. Using the $41\arcsec\times36\arcsec$-resolution image (Fig. \ref{fig:lres}) we estimated the integrated flux over an area of radius of $3\arcmin-5\arcmin$ (Fig. \ref{fig:lres}) is not higher than $25-69$ mJy at 143 MHz at $1\sigma_{\text{\tiny rms}}$ significance.

\section{Discussion}
\label{sec:discussion}

\subsection{Radio relics}
\label{disc_radio_relics}

\cite{Bonafede2009} discussed possible formation models for the radio relics in A1240. The models were associated with large-scale outward propagating shocks generated during the cluster merger and included (i) shock acceleration via Fermi-I process \citep{Ensslin1998,Roettiger1999a,Hoeft2007} and (ii) shock re-acceleration of fossil plasma via adiabatic compression \citep{EnBlin2001}. Using our high-resolution, large-frequency range, and deep LOFAR, GMRT and VLA data, we discuss below the implications of our observational results (Sec. \ref{sec:results}) in the framework of the relic formation models.

\subsubsection{Injection spectral index and shock Mach number}
\label{sec:dis_spx_inj}

The predictions of particle \mbox{(re-)acceleration}
 models at shock fronts depends on the Mach number of shocks \citep[e.g.][]{Donnert2016,Kang2016b} that is defined as follows,

\begin{equation}
	\mathcal{M}=\frac{v_\text{\tiny shock}}{c_\text{\tiny s}},
	\label{eq:M}
\end{equation}
where $v_\text{\tiny shock}$ is the shock speed and $c_\text{\tiny s}$ is the sound speed in the upstream ICM. For simple planar shocks in the linear test particle regime of DSA, the spectral index of the relativistic electrons that are injected at the shock front is a function of the Mach number \citep{Blandford1987},

\begin{equation}
	\alpha_\text{\tiny inj} = \frac{1}{2}-\frac{\mathcal{M}^2+1}{\mathcal{M}^2-1} \quad \text{or} \quad
	\mathcal{M}=\sqrt{\frac{2\alpha_{\text{\tiny inj}}-3}{2\alpha_{\text{\tiny inj}}+1}} . 
\label{eq:spx_inj}
\end{equation}
where the injection spectral index $\alpha_\text{\tiny inj}=(1-	\delta_\text{\tiny inj})/2$, here $\delta_\text{\tiny inj}$ is the power of the particle power spectrum, $dN/dE\propto E^{-\delta_\text{\tiny inj}}$. 

The injection spectral index for a simple planar shock model has been commonly estimated in the literature by using an approximation,

\begin{equation}
	\alpha_\text{\tiny inj} = \alpha_\text{\tiny int} + \frac{1}{2},
\label{eq:spx_inj_int}
\end{equation} 
where $\alpha_\text{\tiny int}$ is the integrated spectral index of the relic.
The advantage of this approach is that the measurement bias is free from the projection and synthesized beam effects as the integrated fluxes are measured over the whole region of the relic. However, in many clusters the Mach numbers derived from the integrated spectral index are higher than those estimated from X-ray data \citep[e.g.][]{Stroe2013a,Akamatsu2015,Eckert2016}. Furthermore, hydrodynamical simulations of cluster shocks indicate that the $\alpha_{\text{\tiny inj}}-\alpha_{\text{\tiny int}}$ approximation (Eq. \ref{eq:spx_inj_int}) does not hold for spherical expanding shocks as the shock speed decreases in time \citep{Kang2015a,Kang2015b}. A second method to estimate the injection spectral index is to directly measure at the shock front with sufficiently high-resolution spectral index maps \citep[e.g.][]{DeGasperin2015,VanWeeren2016b,Hoang2017a}. It is noted that this direct measurement of the injection spectral index is only applicable to the shocks that are moving on/close the plane of the sky to minimize the mixing of different aged electrons. A third method to estimate the injection spectral index is to model the spectral ageing of the relics \citep{Harwood2013, Harwood2015,Stroe2014g,DeGasperin2015}. The spectral ageing modelling requires observations at least 4 frequencies to constrain the spectral curvature of the relics, which we are unable to do with our current data sets. The estimation of injection spectral indices for radio relics using the three approaches above have pros and cons which were discussed in the literature \cite[e.g.][]{Stroe2014g,DeGasperin2015,Hoang2017a}.

In case of A1240, we estimated the integrated spectral indices between 143 MHz and 3 GHz to be $-1.08\pm0.05$ and $-1.13\pm0.05$ for A1240-1 and A1240-2, respectively (Subsec. \ref{sec:res_spx_analysis}). Using Eqs. \ref{eq:spx_inj} and \ref{eq:spx_inj_int}, we estimated the injection spectral indices and the corresponding Mach numbers for the relics to be $-0.58\pm0.05$ and $5.1^{+3.1}_{-1.1}$ for A1240-1 and $-0.63\pm0.05$ and $4.0^{+1.1}_{-0.6}$ for A1240-2. These Mach numbers are significantly higher than those (i.e. $3.3\pm0.2$ and $2.8\pm0.3$, respectively) that were estimated with the VLA 325 MHz and 1.4 GHz data using the above approach reported in \cite{Bonafede2009}. Using the second approach of measuring the injection spectral index directly at the shock front (Subsec. \ref{sec:res_spx_analysis}), we obtained injection spectral indices of $-0.94\pm0.06$ and $-0.97\pm0.05$ for the A1240-1 and A1240-2 shocks, respectively. The corresponding Mach numbers are $2.4\pm0.1$ and $2.3\pm0.1$. The Mach numbers we have estimated are significantly different from each other. A possible reason for this discrepancy, as pointed out in \cite{Kang2015a}, is that the shock compression ratio and the flux of the injected relativistic electrons reduce as the shock speed decreases in time. These lead to a significant deviation of the integrated spectra of the relics from the power laws of the simple planar shock model which predicts the $\alpha_{\text{\tiny inj}}-\alpha_{\text{\tiny int}}$ relation (Eq. \ref{eq:spx_inj_int}). Therefore, the integrated spectra of the relics may be an inaccurate proxy for Mach numbers. However, the spectra of the relics at the location of the shock fronts are properly described by the DSA predictions \citep{Kang2015a} and should be used for the estimates of Mach numbers.

\begin{figure*}
	\centering
	\includegraphics[width=0.8\textwidth]{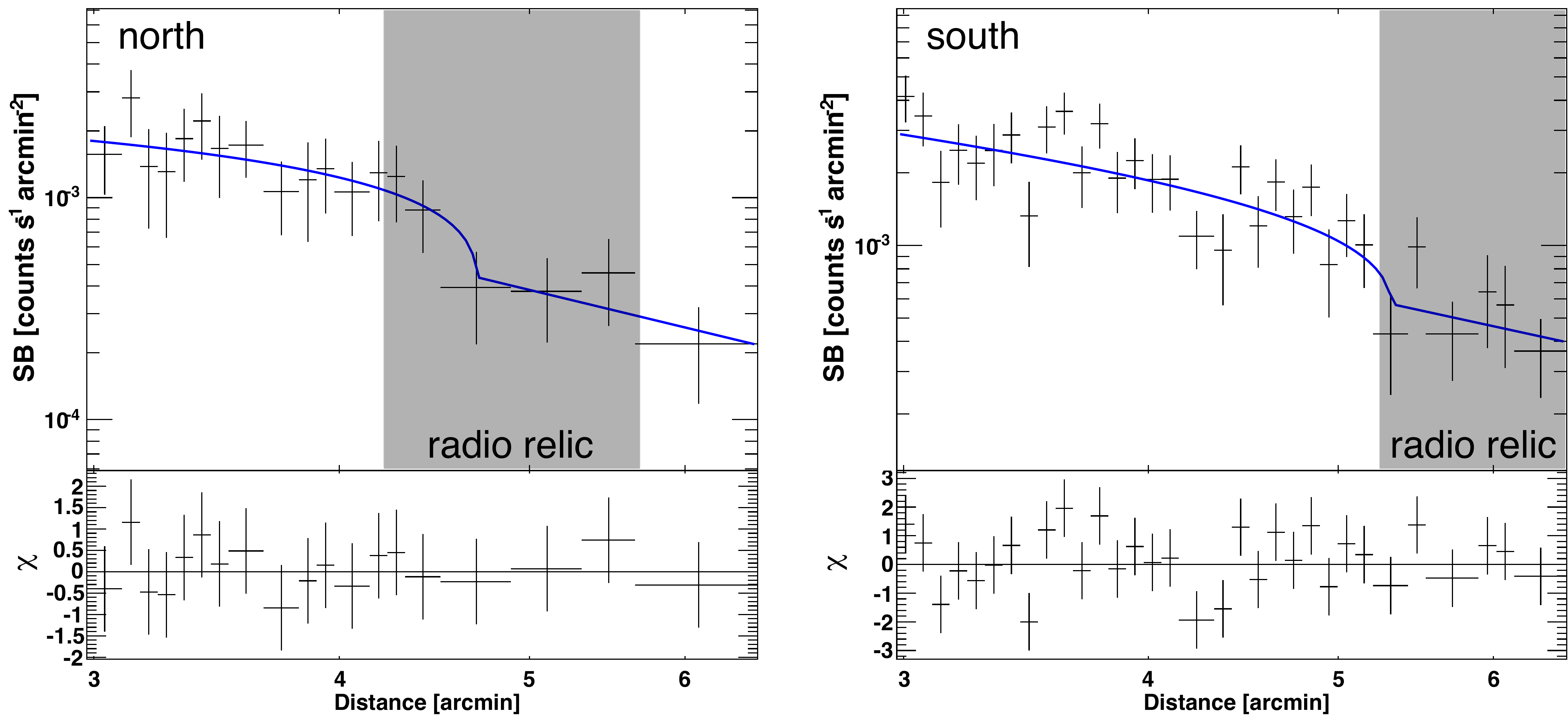} \\
	\caption{52 ks Chandra $0.5-2.0$ keV surface brightness profiles across A1240-1 (left) and A1240-2 (right). The blue lines are the fit of the data to a function consisting of a $\beta$-model and a power law.}
	\label{fig:SB_profile}
\end{figure*}

We analysed \textit{Chandra} X-ray data to search for shocks at the relic locations. We fit the surface brightness (SB) with a function consisting of a $\beta$-model and a power law using $\mathtt{PROFFIT}$ \citep[see, e.g.,][]{Eckert2011,Andrade-Santos2016}. The X-ray SB profiles in Fig. \ref{fig:SB_profile} indicates possible discontinuities across A1240-1 and A1240-2 which would imply the presence of shocks or cold fronts at the location of the relics. To distinguish the nature of the possible discontinuities, a more detailed temperature map is required, which is not possible to make with the current shallow X-ray data. In Fig. \ref{fig:SB_profile}, the SB profile has a break close to the central location of A1240-1 and a SB discontinuity towards the southern direction is found at the inner region of A1240-2. If the relics trace the candidate merger shocks, the locations of these breaks seem to be inconsistent with the DSA model that requires shock fronts to be located at the flattest  spectral regions (i.e. the outer regions) of the relics which is where the relativistic electrons are \mbox{(re-)accelerated}  \citep[e.g.][]{Ensslin1998,VanWeeren2010a}. However, it is known that positional shifts between the relic and X-ray shock positions can occur due to the contamination of small-scale substructure behind the shock that is unresolved, by low-resolution X-ray  observations \citep{Ogrean2013c,VanWeeren2016b}, or due to the contamination of foreground X-ray emission in hydrodynamical simulations or due to projection effects \citep{Hong2015}.  Finally, it is noted that the X-ray data is very shallow (i.e. exposure duration of 52 ks) and the apparent location of the shock fronts in Fig. \ref{fig:SB_profile} might be biased by the low S/N of the X-ray data. 

In an attempt to obtain approximate estimates of the shock Mach numbers with the current X-ray data, we find that the best-fit density jumps would imply Mach numbers of $\sim2$ for both relics, assuming that the density jumps trace two shock fronts. These Mach numbers are in line with our estimates using the radio data (i.e. $\mathcal{M}_\text{\tiny inj}$ in Table \ref{tab:Int_Spx_Mach}). However, future X-ray studies with deeper X-ray/SZ observations will be necessary to provide accurate constraints on the Mach numbers and the exact locations of the shock fronts.

\subsubsection{Acceleration efficiency and sources of relativistic electrons}
\label{sec:disc_electron_sources}

A number of radio relics have been observed at the locations of merger shocks detected with X-ray observations (e.g. via surface brightness discontinuity and/or temperature jump). The shocks are generally thought to accelerate the ICM electrons to relativistic energies and are visible in the radio band under the presence of the large-scale, $\mu$G cluster magnetic field. The Mach numbers for the merger shocks are typically measured to be $\lesssim3$ from X-ray observations \citep[e.g.][]{Markevitch2010,Akamatsu2013}. For these weak shocks, the efficiency to accelerate electrons to relativistic energies directly from the thermal pool can be challenging in the framework of DSA theory \citep[e.g.][]{Kang2012,Pinzke2013,Brunetti2014,Botteon2016,Eckert2016,VanWeeren2016b}. Here the particle acceleration efficiency is defined as follows \citep{Botteon2016},

\begin{equation}
	\eta = \frac{E_\text{\tiny relic}}{\Delta F_\text{\tiny KE}},
	\label{eq:PAE}
\end{equation}
where $E_\text{\tiny relic}$ is the energy flux of the accelerated relativistic electrons at relic and $\Delta F_\text{\tiny KE}$ is the kinematic energy flux available at the shock,  

\begin{equation}
E_\text{\tiny relic}=\epsilon_\text{\tiny e,down} v_\text{\tiny down}
\label{eq:E_relic}
\end{equation}

\begin{equation}
	\Delta F_\text{\tiny KE}=0.5\rho_\text{\tiny up}v^3_\text{\tiny shock}(1-\frac{1}{C^2}),
	\label{eq:DeltaF}
\end{equation}
where $\epsilon_\text{\tiny e,down}$ and $v_\text{\tiny down}$ are the downstream particle energy density and velocity, respectively; $\rho_\text{\tiny up}$ is the upstream density; $v_\text{\tiny shock}$ is the shock speed; $C=\frac{(\gamma+1)\mathcal{M}^2}{(\gamma-1)\mathcal{M}^2+2}$ is the compression factor of a shock Mach number $\mathcal{M}$ (here $\gamma=5/3$). The relativistic electrons in the downstream region were assumed to have a single-power law spectrum, $N_\text{\tiny inj}\propto p^{-\delta_\text{\tiny inj}}$. For details of the formulas, we refer to \cite{Botteon2016}.

\begin{figure}
	\centering
	\includegraphics[width=1\columnwidth]{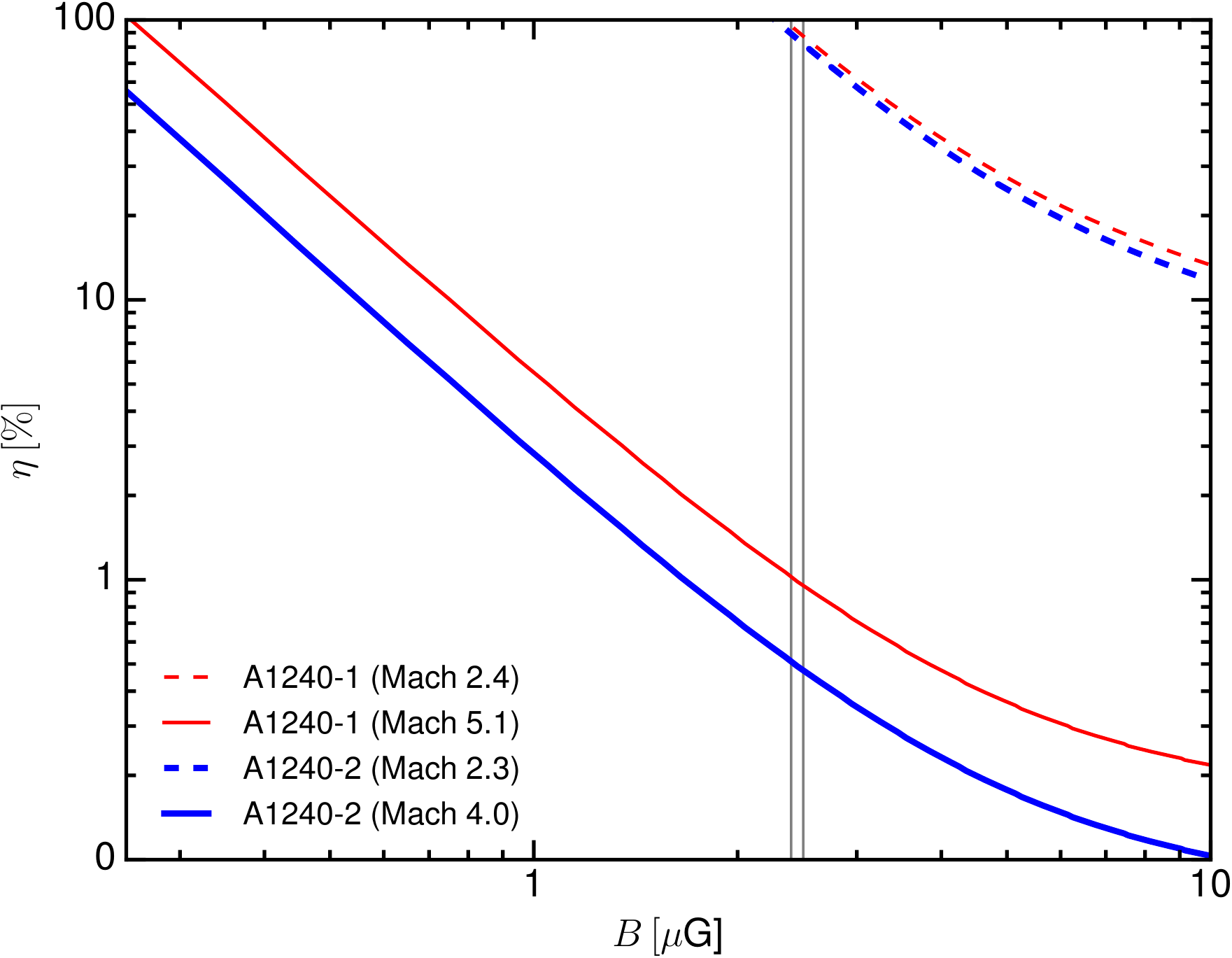} \\
	\caption{Particle acceleration efficiency $\eta(B)$ required to produce the radio surface brightness in the relics of A1240. The vertical lines indicate the equipartition magnetic field $2.4\,\mu\text{G}$ and $2.5\,\mu\text{G}$ in A1240-1 and A1240-2, respectively \citep{Bonafede2009}. Calculations assume a minimum momentum of electrons $p_\text{\tiny min}=0.1 m_e c$.}
	\label{fig:PA_eff}
\end{figure}

In Fig. \ref{fig:PA_eff} we examine the particle acceleration efficiency for shocks with the injection indices (or Mach numbers) for the relics A1240-1 and A1240-2 (see Table \ref{tab:Int_Spx_Mach}). In the calculation, we used the downstream particle number densities $\rho_\text{\tiny A1240-1}=(1-2)\times10^{-4}\,\text{cm}^{-3}$ and $\rho_\text{\tiny A1240-2}=(2.5-3.5)\times10^{-4}\,\text{cm}^{-3}$, which were derived by fitting of the electron density beta-model profile to the \textit{Chandra} X-ray data. We also used the downstream temperature $T_\text{\tiny A1240-1}=5.1^{+1.0}_{-0.8}\,\text{keV}$ and $T_\text{\tiny A1240-2}=5.4^{+0.9}_{-0.8}\,\text{keV}$ \citep{Barrena2009}. The $k$-corrected radio power used for the relics in the calculation is $P_\text{\tiny A1240-1}^\text{\tiny 143 MHz} = (7.52\pm0.17)\times 10^{24}\,\text{W\,Hz}^{-1}$ and $P_\text{\tiny A1240-2}^\text{\tiny 143 MHz} = (2.24\pm0.33)\times10^{25}\,\text{W\,Hz}^{-1}$ that we calculated from the LOFAR image (see Table \ref{tab:int_fluxes}). Given the equipartition magnetic field strength of $\sim2.5\,\mu\text{G}$ in the relic regions \citep{Bonafede2009}, in the cases of higher Mach numbers (i.e. 4.0 for A1240-1 and 5.1 for A1240-2) the particle acceleration efficiencies that are required to produce the synchrotron emission in A1240-1 and A1240-2 are less than $1\%$ and $0.5\%$, respectively. Although the precise efficiency of electron acceleration by the low Mach numbers of shocks associated with the relics is still an open question, these low efficiencies are likely to be realistic  \citep{Brunetti2014}. However, the required efficiencies for low Mach numbers (e.g. $\lesssim3$) are close to 100 percent which is challenging for DSA. To avoid the high efficiency problem, it is proposed that the low Mach number shocks re-accelerate a pre-existing population of relativistic electrons, instead of accelerating the thermal electrons \citep[e.g.][]{Markevitch2005,Kang2011a,Kang2012}. The pre-existing fossil plasma could originate from radio galaxies that are close to the relics. To search for sources of possible fossil plasma, we obtained the redshifts from the Subaru and SDSS optical data \citep{Golovich2017a} for the radio galaxies (i.e. A, B, C, E, F in Fig. \ref{fig:hres_vla_gmrt}) that have small angular separations to A1240-1 and A1240-2. The galaxies C ($z=0.888\pm0.0979$) and E ($z=0.448\pm0.0289$) are background sources and D has no redshift information. The radio galaxies A ($z=0.19299\pm0.00003$), B ($z=0.19223\pm0.00005$), and possibly F ($z=0.152\pm0.0263$) are close to the cluster mean redshift ($z=0.1948$) and are possibly sources of mildly relativistic electrons that could be associated with the synchrotron radio emission in the relics. An example of this scenario was observed in Abell 3411-3412 where fossil electrons from a radio galaxy have been suggested to be re-accelerated by a merger shock which disturbs the morphology of the tails at the location of the shock and re-flattens the spectral index of the tails at the location of the shock \citep{VanWeeren2017}. Other less obvious examples are found in PLCKG287.0+32.9 \citep{Bonafede2014} and the Bullet cluster 1E 0657-55.8 \citep{shimwell2015}. As our radio data presented in Figs. \ref{fig:hres_lofar} and \ref{fig:hres_vla_gmrt} are not deep enough to provide information on whether A, B and F are connected to A1240-1 and A1240-2 and do not allow us to study the spectral index trend of the sources, future deeper, high-resolution radio observations will be necessary to establish such a connection.

\subsubsection{Size and power of the double relics}
\label{sec:lls}

In the DSA model, the extent of radio relics is the same as the size of the shock fronts which (re-)accelerate in situ the relativistic electrons in the relics \cite[e.g.][]{Jaffe1977,Blandford1987,Ensslin1998}. In merging clusters that host double radio relics on opposite sides of the cluster centre, the relative largest linear size (LLS) of the relics depends on the mass ratio of the sub-clusters, as demonstrated in, e.g., hydrodynamical simulations of ideal binary cluster merger by \cite{VanWeeren2011}. In these simulations, the sub-cluster mass ratio is varied to match the observed LLSs of double relics (i.e. in the Sausage cluster). The larger relic is found to be behind the more massive sub-cluster. Although the mass configuration (i.e. ratio of $1-3:1$) in the simulations might be inconsistent with the reconstructed mass distribution in the weak lensing data (i.e. mass ratio $\sim1$ in \citealt{Jee2015} or $\sim1:2$ in \citealt{Okabe2015a}), this might be because the observed LLS of the faint, steep spectrum relic were biased low by the sensitivity limitations of the high-frequency radio observations used in \cite{VanWeeren2011}. We note that the error bars associated with the weak lensing analyses are so large that it is not clear there is real discrepancy. In support of this, it is known that the LLS of the small relic is much large and almost equal to the main relic \citep{Hoang2017a}. It is also noted that the observed LLS of a relic also depends on mass concentration of the sub-clusters prior to merger. In line with the simulations in \cite{VanWeeren2011}, a number of merging clusters are observed to host more massive sub-clusters behind the main relics (e.g. ZwCl 0008.8+5215 in \citealt{VanWeeren2011a}, \citealt{Golovich2017a}; RX J0603.3+4214 in \citealt{VanWeeren2016b}, \citealt{Jee2016}; and PLCK G287.0+32.9 in \citealt{Bonafede2014}, \citealt{Finner2017}).

In A1240, the projected LLS of A1240-2 in the south is a factor of 2 larger than that of A1240-1 in the north (i.e.  $1.35$ and $0.68\,\text{Mpc}$, respectively; see Subsec. \ref{sec:res_radio_relics}). This implies that the southern shock front is larger in projection than the north one. Unfortunately, we are unable to check this with the current shallow X-ray data. However, if this turns out to be the case and the mass concentration of the sub-clusters is similar, the sizes of the shock fronts is likely different because the masses of the sub-clusters prior to the merger are not equal. The sub-cluster that is more massive (or larger in size) should generate a larger shock surface in front of its direction of propagation during a major cluster merger. This scenario might be applied for A1240 as the cluster is known to be observed $0.3\,\text{Gyr}$ after core crossed \citep{Barrena2009}. The southern sub-cluster, which comes from the north before the merger, is more massive (about 2.8 times; \citealt{Barrena2009}) than the northern sub-cluster and generates a wider shock in the south than the northern counter shock.

The radio power at 143 MHz for A1240-2 is a factor of $\sim3$ more powerful than that for A1240-1 (see Subsec. \ref{sec:disc_electron_sources}). This is surprising because the radio derived Mach number for A1240-2 is smaller or equal to that for A1240-1 (Table \ref{tab:Int_Spx_Mach}). This might be because the surface area in the southern shock is larger than that in the northern shock which might be due to the difference in the mass of the sub-clusters, as we discussed above. The true reason is still unclear as the synchrotron power of the relics under DSA model is a function of many parameters (e.g. shock surface area, electron density, magnetic field strengths, ICM temperature, particle acceleration efficiency at the shocks; Eq. 32 in \citealt{Hoeft2007}) that are poorly constrained with the current data. Other possibilities for the difference in the power of the relics are that the relativistic electrons in the relics are re-accelerated from fossil plasma and the radio power depends on the fossil plasma populations \citep[e.g.][]{VanWeeren2016b} or the Mach numbers derived from radio spectrum  are not an approximate proxy for the X-ray shock Mach numbers \citep[e.g.][]{Akamatsu2015,Akamatsu2017,VanWeeren2016b}.

\subsubsection{Viewing angle of the merger axis}
\label{sec:disc_viewing_angle}

The radio emission from the relics A1240-1 and A1240-2 (Fig. \ref{fig:pols_F}) is highly polarized. The electric field vectors are roughly perpendicular to the major axes of the relics which implies an alignment of magnetic fields along the major axes of the relics. At the shock fronts, the magnetic field alignments are likely to be caused by shock compression. Since the polarized emission is a vector quantity, the fractional polarization as measured by an observer depends on the viewing angle  $\theta$ and the compression factor $C=(\alpha_\text{\tiny int}-1)/(\alpha_\text{\tiny int}+\frac{1}{2})$ (assuming a polytropic index of the ICM gas of $\gamma_\text{\tiny gas}=5/3$, \citealt{Ensslin1998}). The viewing angle here is the projection angle between the normal of the shock front and the line from the observers to the shock; for example, $\theta=0^\circ$ or $90^\circ$ means that the shock is occurring along the LOS or in the plane of the sky, respectively. In cases of a weak magnetic field or small ratio of the magnetic pressure to internal gas pressure (i.e. $P_\textit{\tiny B}/P_\text{\tiny thermal}=\frac{B^2}{8\pi \rho R T}$, where $B$ is magnetic field, $\rho$ is thermal electron density, $R$ is ideal gas constant, $T$ is thermal gas temperature), the observed mean fractional polarization of a shock is estimated as following \citep{Ensslin1998},

\begin{equation}
\bar{F} \leqslant \frac{\delta_\text{\tiny int}+1}{\delta_\text{\tiny int}+\frac{7}{3}}\frac{\sin^2(\theta)}{\frac{2C^2}{C^2-1}-\sin^2(\theta)},
\label{eq:F_angle}
\end{equation}
where $\delta_\text{\tiny int}=1-2\alpha_\text{\tiny int}$ is the slope of the electron density spectrum. The $\leqslant$ sign indicates that the observed polarized emission might further experience depolarization effects due to, e.g., the spatial resolution of the observations or wide bandwidth imaging. In case of A1240, we estimated the magnetic field to thermal pressure ratios of $\sim18\%$ for A1240-1 and $\sim9\%$ for A1240-2, which implies that the relics are located in a region that satisfies the weak magnetic field criteria. Here we used the magnetic field strength (i.e. $B_\text{\tiny A1240-1}=2.4\,\mu \text{G}$, $B_\text{\tiny A1240-2}=2.5\,\mu \text{G}$) estimated in \cite{Bonafede2009}, the particle upstream densities (i.e. $\rho_\text{\tiny A1240-1}=1.5\times10^{-4}\,\text{cm}^{-3}$, $\rho_\text{\tiny A1240-2}=3.0\times10^{-4}\,\text{cm}^{-3}$) calculated in Sec. \ref{sec:disc_electron_sources} and the thermal temperature (i.e. $T_\text{\tiny A1240-1}=5.1\,\text{keV}$ and $T_\text{\tiny A1240-2}=5.4\,\text{keV}$) measured in \cite{Barrena2009}.

\begin{figure}
	\centering
	\includegraphics[width=1\linewidth]{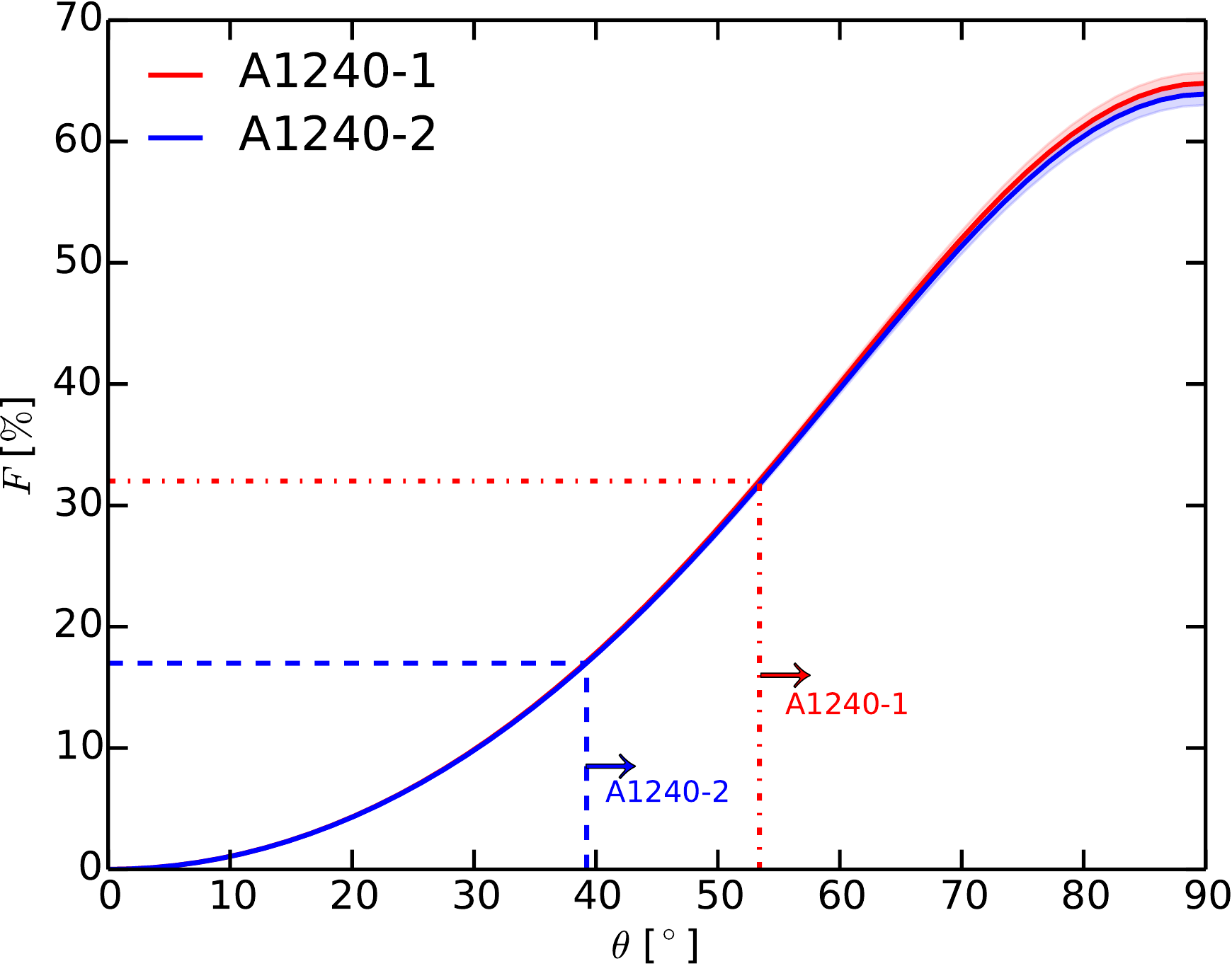} \\
	\caption{Fractional polarization of the relics A1240-1 and A1240-2 as a function of viewing angle. The mean fractional polarization of A1240-1 and A1240-2 was measured to be $32\pm4\%$ and $17\pm4\%$ from the $480\,\text{MHz}$-bandwidth data chunks (Subsec. \ref{sec:res_pol_analysis}), which correspond to viewing angles of $\theta_\text{\tiny A1240-1}\geqslant53\pm3^{\circ}$ (red arrow) and $\theta_\text{\tiny A1240-2}\geqslant39\pm5^{\circ}$ (blue arrow), respectively.}
	\label{fig:pols_angle}
\end{figure}

To examine the possible viewing angles of the relics A1240-1 and A1240-2, we plot the fractional polarization $\bar{F}$ of the relics as a function of viewing angle $\theta$ in Fig. \ref{fig:pols_angle}.  Here we used the integrated spectral indices of $-1.08\pm0.05$ and $-1.13\pm0.05$ for A1240-1 and A1240-2, respectively (Subsec. \ref{sec:res_spx_analysis}). Since the polarization measured from the VLA $2-4$ GHz data might be slightly depolarized due to the wide-bandwidth, we used the mean fractional polarization that was measured from the VLA images (bandwidth of 480 MHz, see Subsec. \ref{sec:res_pol_analysis}). The mean fractional polarization measured from the VLA data sets are $32\pm4\%$ and $17\pm4\%$ for A1240-1 and A1240-2, respectively. These correspond to the viewing angles of $\theta_\text{\tiny A1240-1}\geqslant53\pm3^\circ$ and $\theta_\text{\tiny A1240-2}\geqslant39\pm5^\circ$ (i.e. via Eq. \ref{eq:F_angle}). The estimated viewing angles are in agreement with the constraints from the two-body modelling using optical redshift data that the cluster merger likely occurred in the plane of the sky \citep{Barrena2009}.

\subsection{Radio halo and cluster mass}
\label{sec:dics_radio_halo}

\cite{Cassano2013a} reports the relation between the power of radio haloes and the cluster mass (i.e. the $P_\text{\tiny 1.4 GHz}-M_\text{\tiny 500}$ relation). The power of radio haloes increases as a function of cluster mass, implying that more gravitational energy is channelled into the radio emission in more massive merging clusters. A number of low-mass ($\sim10^{14}M_{\odot}$) merging clusters have also been detected to host radio haloes. However, diffuse emission has not been observed in the central regions of some clusters in a similar mass range (e.g. Abell 2345 in  \citealt{Bonafede2009}, ZwCl 2341.1+0000 in \citealt{VanWeeren2009b}, ZwCl 0008.8+5215 in \citealt{VanWeeren2011a}, PSZ1 G096.89+24.17 in \citealt{deGrasperin2014}). The question remains as to what fraction of merging galaxy clusters are not able to generate radio haloes or whether the non-detection is simply due to the sensitivity limitation of the current radio observations. In case of A1240, our LOFAR 143 MHz data (Fig. \ref{fig:lres}) shows no evidence of diffuse emission in the central region of the cluster despite of the fact that its ICM is highly disturbed. In this subsection, we use the LOFAR data to constrain the upper limit for the flux of the radio halo. We assume that the spectral index for the halo is $\alpha_{\text{\tiny int}}=1.3\pm0.2$ (i.e. the typical spectral index for a number of known haloes; \citealt{Luigina2012}) and the radio power is $k$-corrected for the redshift of the cluster.

\subsubsection{Upper limit for halo flux from the LOFAR data}
\label{sec:halo_mass}

Given the cluster mass of $M_\text{\tiny 500}=(3.7\pm0.4)\times10^{14}\,M_\odot$  \citep{Planck2015}, we found the predicted power for A1240 to be $P_\text{\tiny 1.4 GHz}=(1.9\pm0.9)\times10^{23}\,\text{W\,Hz}^{-1}$. The power of radio haloes is also found to be directly proportional to their radii, $R_\text{\tiny H}=\sqrt{R_\text{\tiny min}\times R_\text{\tiny max}}$, where $R_\text{\tiny min}$ and $R_\text{\tiny max}$ are the minimum and maximum radii of the haloes \citep[e.g.][]{Cassano2007}. Using the $P_\text{\tiny 1.4 GHz}-R_\text{\tiny H}$ relation in \cite{Cassano2007}, we estimate that the radius for the halo in A1240 is $R_\text{\tiny H}=264\,\text{kpc}$, which is significantly smaller than half of the distance between A1240-1 and A1240-2 (i.e. $900\,\text{kpc}$). 

The $1\sigma_{\text{\tiny rms}}$ upper limit for the flux of non-detected diffuse emission is commonly estimated as $S=A\times\sigma_{\text{\tiny rms}}$, where $A$ is the area emitting the expected diffuse emission. Given the sensitivity of the LOFAR data (Fig. \ref{fig:lres}), we estimated that the halo flux at 143 MHz is at most $S_\text{\tiny 143 MHz}= A \times\sigma_{\text{\tiny rms}}=5.1\,\text{mJy}$, where $\sigma_\text{\tiny rms}=410\,\mu\text{Jy}\,\text{beam}^{-1}=0.244\,\mu\text{Jy}\,\text{arcsec}^{-2}$ and $A=\pi R_\text{\tiny H}^2= 21\times10^{3}\,\text{arcsec}^2$, corresponding to a radius of $264\,\text{kpc}$. The derived upper limit for the halo power at 1.4 GHz is $P_\text{\tiny 1.4 GHz}=(3.0\pm1.4)\times10^{22}\,\text{W\,Hz}^{-1}$. Our estimated $1\sigma_{\text{\tiny rms}}$ upper limit for the radio halo is $\sim6-12$ times smaller than the value predicted from the  $P_\text{\tiny 1.4 GHz}-M_\text{\tiny 500}$ relation (i.e. $P_\text{\tiny 1.4 GHz}=(1.9\pm0.9)\times10^{23}\,\text{W\,Hz}^{-1}$; \citealt{Cassano2013a}). The radio halo in A1240 is therefore underluminous in comparison with the prediction of the $P_\text{\tiny 1.4 GHz}-M_\text{\tiny 500}$ relation if it has a small radius (i.e. $R_\text{\tiny H}$). Several other double-relic clusters are observed to host small haloes \citep[e.g.][]{Brown2011a,Bonafede2012,DeGasperin2015}. But it is noted that the apparent sizes of faint radio haloes might be biased due to, e.g., the depth or $uv$-coverage of the observations.

\begin{table}
	\centering{}
	\caption{Upper limits for the radio halo power of A1240.}
	\begin{tabular}{lc}
		\hline\hline
		Region                                    & $P_\text{\tiny 1.4 GHz} [10^{23}\,\text{W\,Hz}^{-1}]$ \\ \hline
		Circle ($R_\text{\tiny H}=264\text{kpc}$) &                     $0.30\pm0.14$                     \\
		Circle (mock injection)                  &             $0.29\pm0.13$              \\
		Ellipse (from Figs. \ref{fig:lres})        &                      $2.2\pm1.0$                      \\

		Ellipse (mock injection)                  &             $1.4\pm0.6$              \\ \hline\hline
	\end{tabular} \\
	\label{tab:upper_limits}
\end{table}

\begin{figure}
	\centering
	\includegraphics[width=1\linewidth]{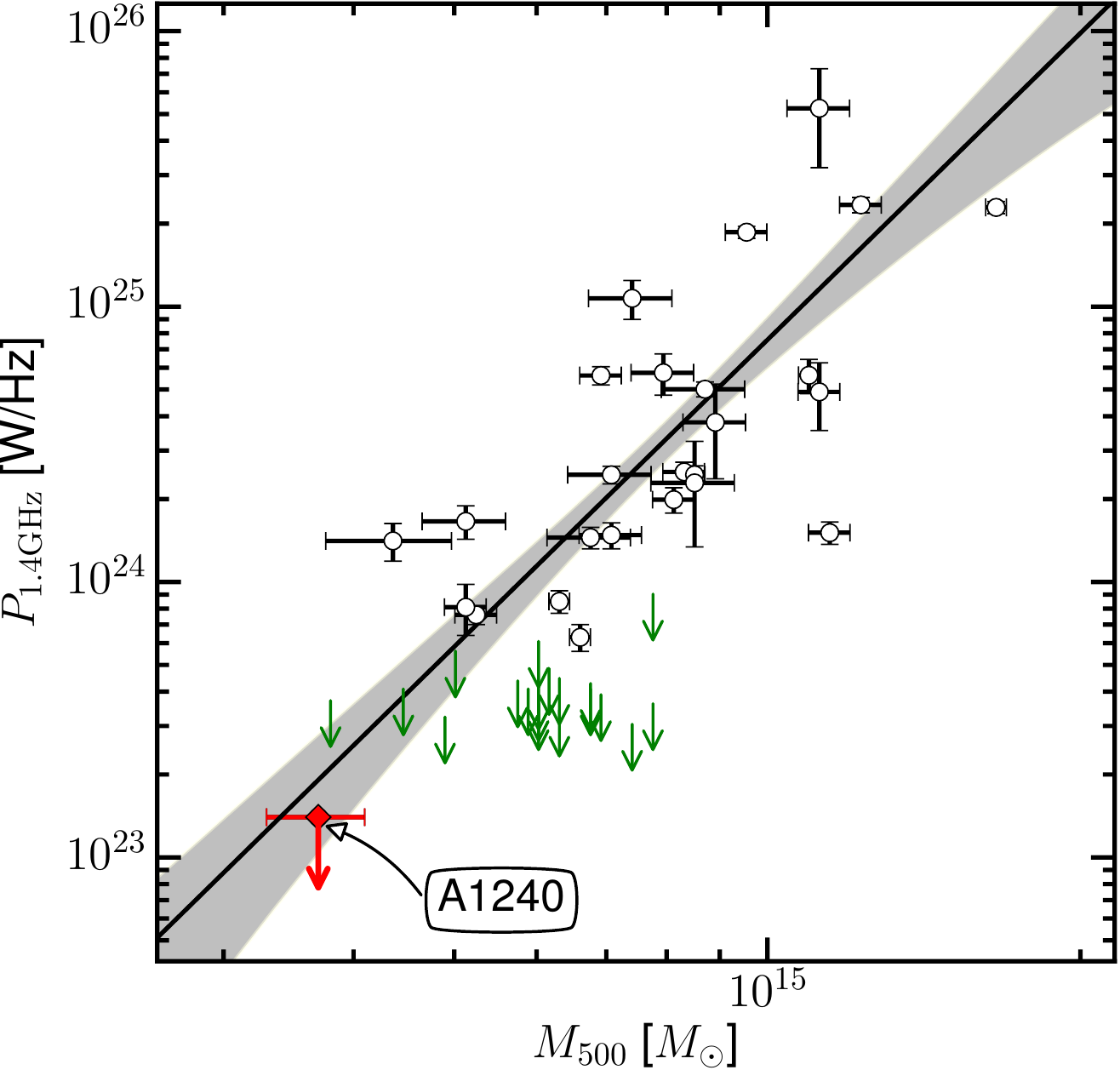}
	\caption{The $P_\text{\tiny 1.4 GHz}-M_\text{\tiny 500}$ relation of radio haloes (i.e. $P_{\text{\tiny 1.4 GHz}} [10^{24}\text{W\,Hz}^{-1}]= 10^{-2.82\pm0.51}\times M_{\text{\tiny 500}}^{3.70\pm0.56}[10^{14}M_{\odot}]$; i.e. BCES bisector fitting for all detected haloes in \citealt{Cassano2013a}). The upper limits of the radio halo for A1240, $P_\text{\tiny 1.4 GHz}=(1.4\pm0.6)\times10^{23}\,\text{W\,Hz}^{-1}$ (red arrow), 
	that we estimated from the LOFAR 143 MHz data in Subsec. \ref{sec:dics_radio_halo}. The green down arrows indicate upper limits for the undetected radio haloes (see  \citealt{Cassano2013a} for the cluster list). The grey region indicate the $95\%$ confidence region of the best-fit.}
	\label{fig:pm500}
\end{figure}

In the post-shock turbulence acceleration scenario, the turbulence is produced by the baroclinic effects at the shocks, where the pressure gradients are not in parallel with the downstream total plasma \citep{Brunetti2014}. Suggested by the Fermi-II process, the radio halo emission may start from the inner edges of the relics, or behind the shocks, and approximately trace the X-ray emission (e.g. \citealt{VanWeeren2016b,Hoang2017a}). If the radio halo of A1240 follows a similar trend, the size of the halo should be approximately equal to the distance between A1240-1 and A1240-2 (i.e. $\sim1.8$ Mpc in projection). If we assume the halo emission encompasses the region shown by the ellipse in Fig. \ref{fig:lres} then the $1\sigma_{\text{\tiny rms}}$ upper limit of the radio flux is significantly higher at $S_\text{\tiny 143 MHz}= A \times\sigma_{\text{\tiny rms}}=37.2\,\text{mJy}$, where $A=152.6\times10^{3}\,\text{arcsec}^2$. The ellipse has semi-minor and semi-major axes of $3$ and $4.5\,\text{arcmin}$, respectively. The position angle of the ellipse is $160^\circ$ (the north-south axis being the reference, the counter-clockwise being positive direction). The corresponding limit on the 1.4 GHz radio power is $P_\text{\tiny 1.4 GHz}=(2.2\pm1.0)\times10^{23}\,\text{W\,Hz}^{-1}$, which is not discrepant from that predicted by the $P_\text{\tiny 1.4 GHz}-M_\text{\tiny 500}$ relation (i.e. $P_\text{\tiny 1.4 GHz}=(1.9\pm0.9)\times10^{23}\,\text{W\,Hz}^{-1}$; see Fig. \ref{fig:pm500}). Hence, even though we do not detect the halo, its expected power should be close to the detection limit of our observations.

\subsubsection{Upper limit of halo flux from simulations}
\label{sec:halo_inj}

Beside the uncertainty in the predictions of the halo size and its radio power, other possibilities for the non-detection of diffuse emission in the cluster centre are due to (i) the limited $uv$-coverage of the interferometric observations \citep[see, e.g., ][]{Venturi2008a} and (ii) the algorithms used in the data reduction procedure. 

\begin{figure}
	\centering
	\includegraphics[width=0.49\linewidth]{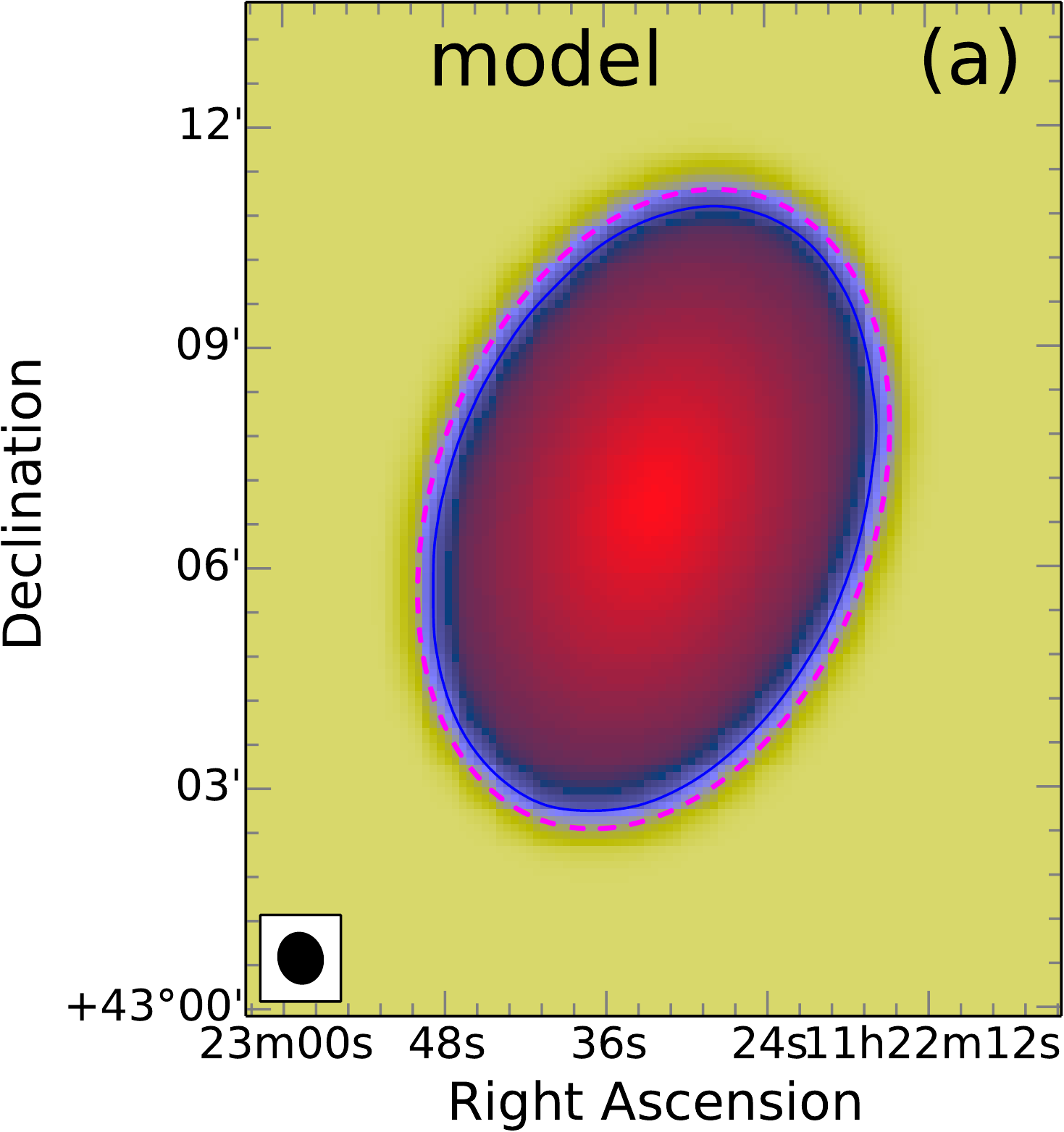} 
	\includegraphics[width=0.49\linewidth]{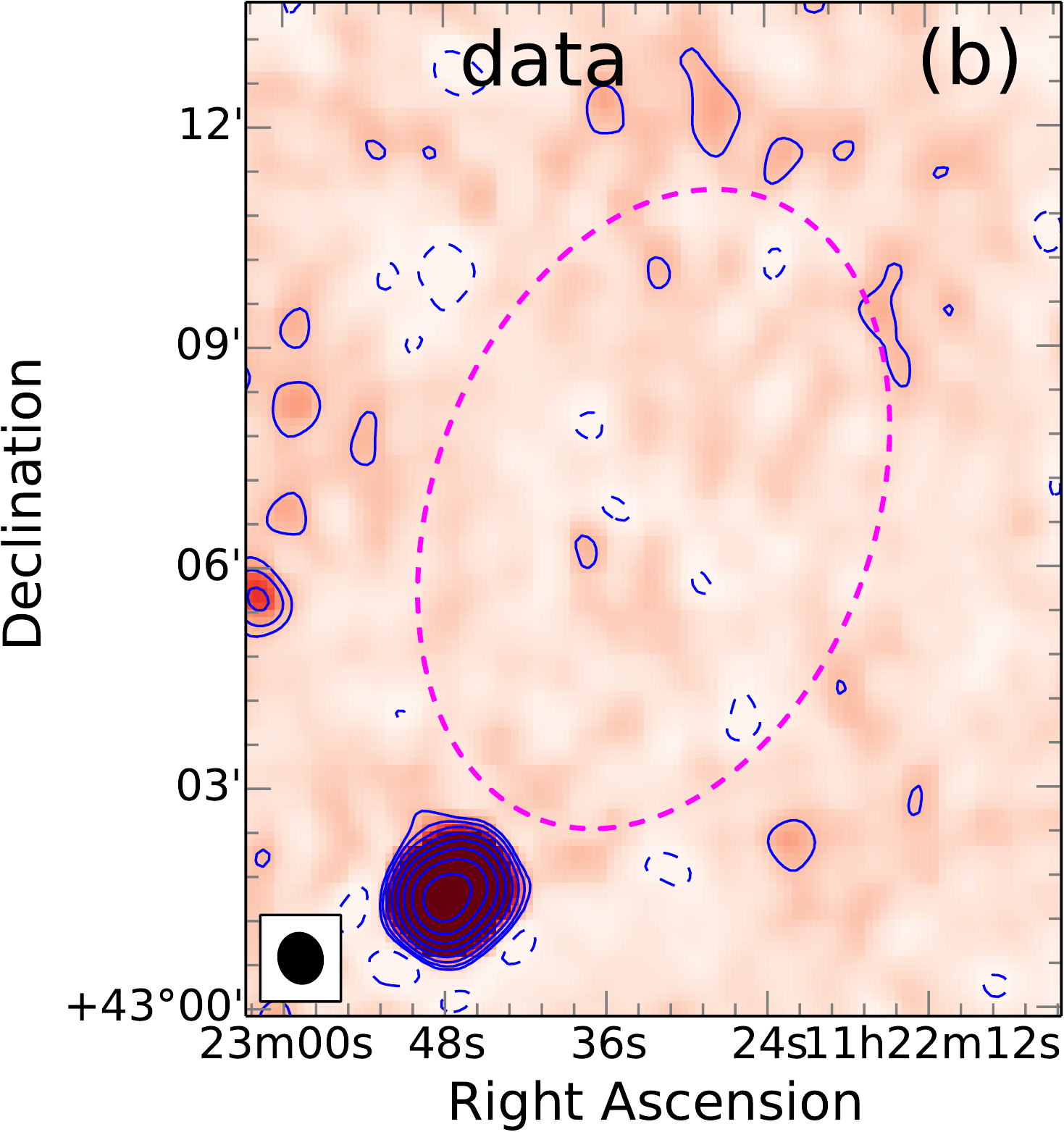} \\
	\includegraphics[width=0.49\linewidth]{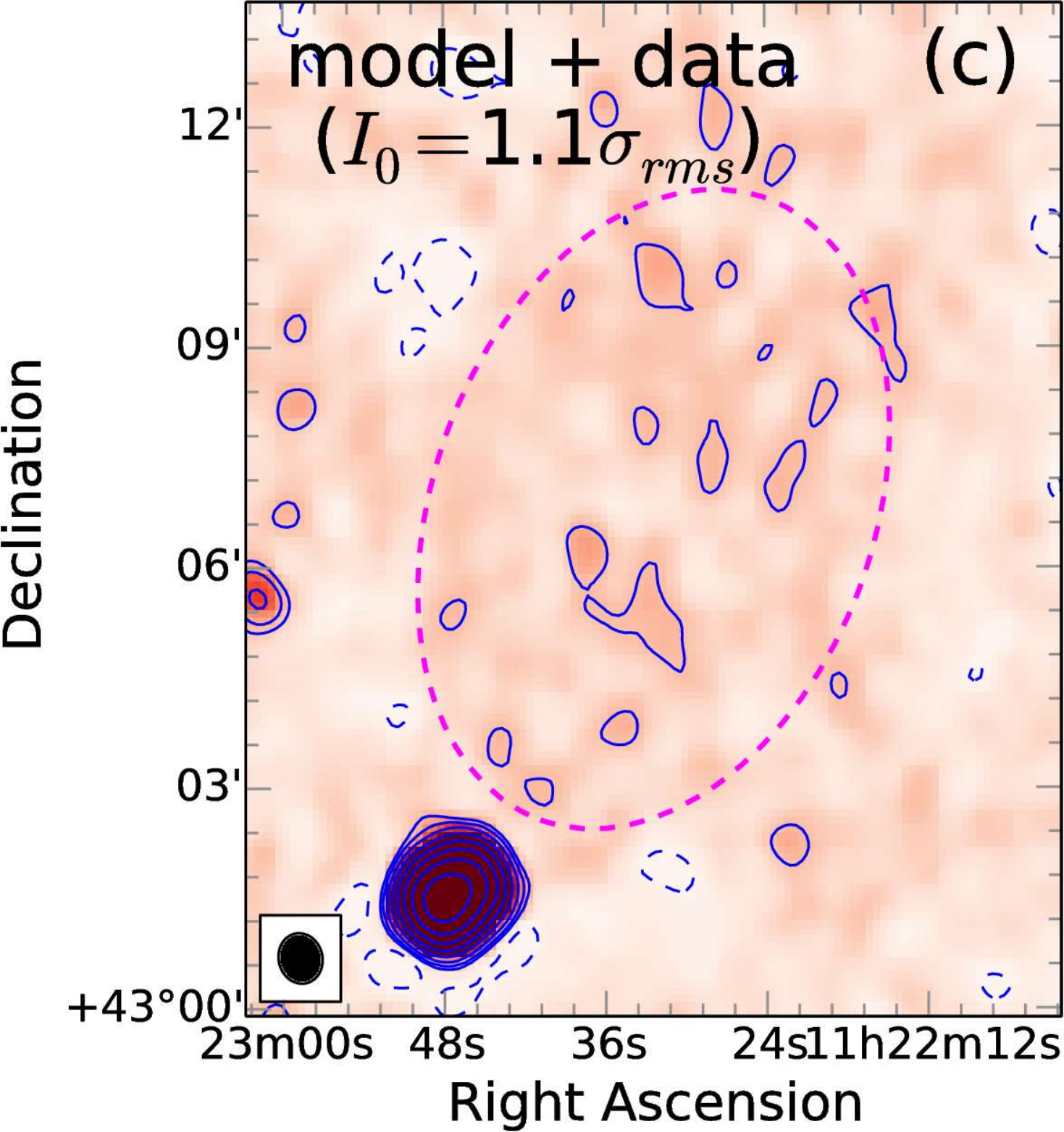} 
	\includegraphics[width=0.49\linewidth]{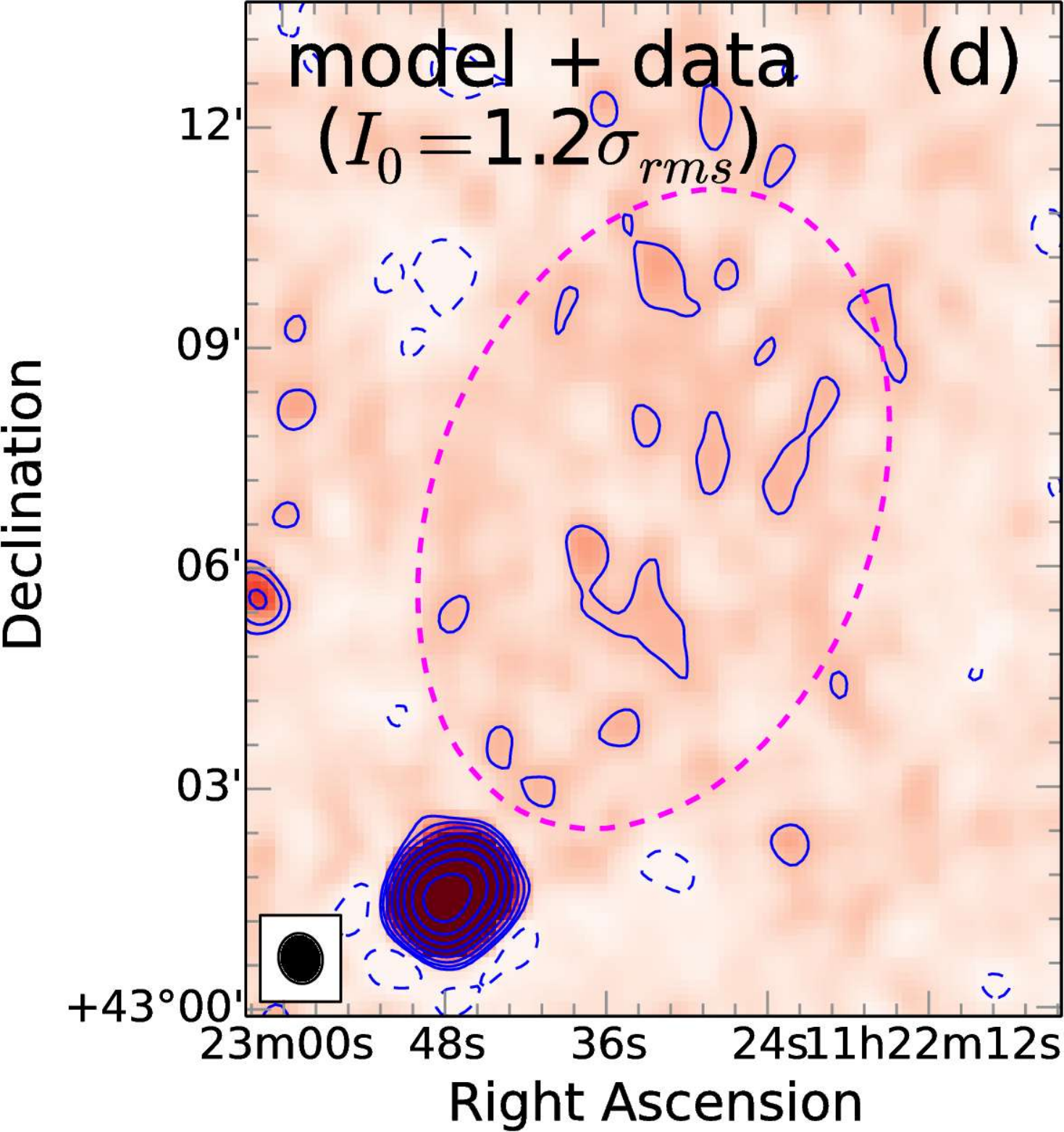} \\
	\includegraphics[width=0.49\linewidth]{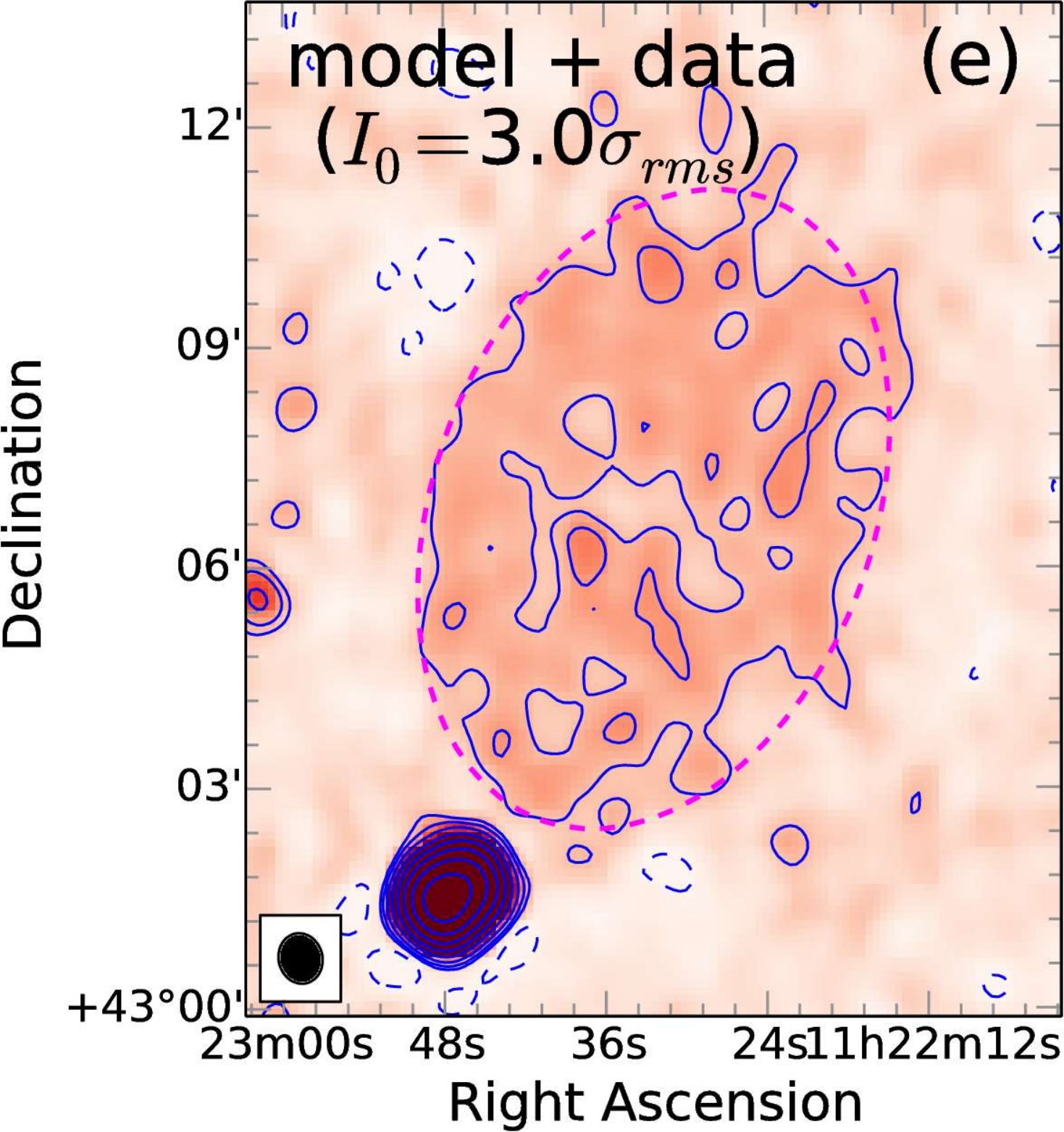} \\

	\caption{Injection of halo models to the LOFAR data. (a) An example of the injected model for the halo (smoothed to $41\arcsec\times36\arcsec$ resolution). (b) LOFAR image of the region where the halo model is injected. (c-e) LOFAR images after the halo models of various peak brightness $I_0$ are injected. The contours start at $\pm2$ and are spaced with a factor of 2. The size of the elliptical region is identical to that in Fig. \ref{fig:lres}.
		}
	\label{fig:halo_inj}
\end{figure}

To examine (i), we  generated a mock image for the radio halo with a surface brightness that follows an exponential function $I(r)=I_\text{\tiny 0} \exp(-\epsilon/r_\text{\tiny e})$, where $\epsilon=\frac{x^2}{a^2}+\frac{y^2}{b^2}\le1$ defines the projected shape of the diffuse emission \citep{Bonafede2017}; $I_\text{\tiny 0}$ and $r_\text{\tiny e}$ are free parameters \citep{Murgia2009}. The mock halo has a peak brightness of $I_\text{\tiny 0}=3\sigma_{\text{\tiny rms}}=0.732\,\mu\text{Jy}\,\text{arcsec}^{-2}$, an effective radius of $r_\text{\tiny e}=102\,\text{kpc}$ and an integrated flux of $100.4\,\text{mJy}$ at 143 MHz over the elliptical region ($a=582\,\text{kpc}$ and $b=873\,\text{kpc}$) in Figs. \ref{fig:lres} and \ref{fig:halo_inj}. The halo model was injected (i.e. Fourier transformed) into a blank column of the LOFAR data set that does not include sources nor background noise. The injected $uv$-data were then deconvolved with the $\mathtt{MS-MFS}$ options in $\mathtt{CASA}$ (see Subsec. \ref{sec:red_contiuum_imaging}) to obtain the mock halo. The integrated flux for the mock halo that was measured in the deconvolved image is $95.9\,\text{mJy}$ which is $\sim 95\%$ of the integrated flux of the injected mock halo. This implies that the LOFAR observations are sensitive to the 143 MHz, $\sim\mu\text{Jy}\,\text{arcsec}^{-2}$ diffuse emission of scales equivalent to the separation of the relics in A1240. 

\begin{figure}
	\centering
	\includegraphics[width=1\linewidth]{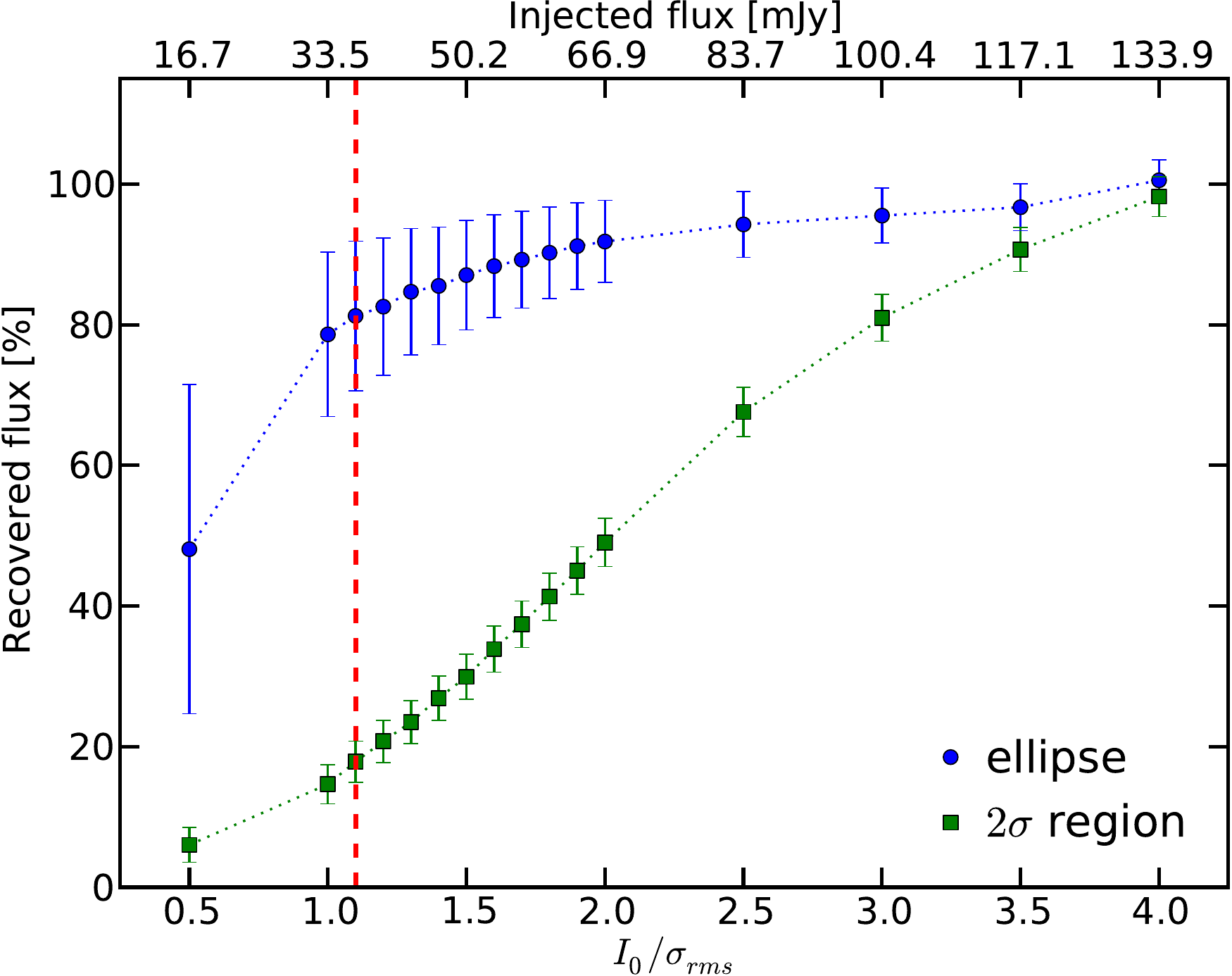}
	\caption{Recovered flux for the radio halo depends on the S/N of the injected elliptical model. The recovered fluxes are measured within the elliptical region and the $\geqslant2\sigma_{\text{\tiny rms}}$ region (see Fig. \ref{fig:halo_inj}). The vertical dashed line indicates the upper limit for the halo flux (i.e. Fig. \ref{fig:halo_inj}, c).}
	\label{fig:flux_recovered}
\end{figure}

To search for the upper limit of the halo flux, we follow the procedure described in \cite{Bonafede2017}. We vary the integrated fluxes (i.e. using $I_\text{\tiny 0}$) and inject the elliptical model ($a=582\,\text{kpc}$,  $b=873\,\text{kpc}$ and $r_\text{\tiny e}=102\,\text{kpc}$) to the LOFAR data column. $I_\text{\tiny 0}$ is varied in steps of $0.5\sigma_{\text{\tiny rms}}$, followed by fine steps of $0.1\sigma_{\text{\tiny rms}}$ between $1-2\sigma_{\text{\tiny rms}}$. We found that no extended emission (i.e. size of $\geqslant R_\text{\tiny H}$) is  detected in the deconvolved images (Fig. \ref{fig:halo_inj}) at $2\sigma_{\text{\tiny rms}}$ when the integrated flux of the injected model is $\leqslant36.8\,\text{mJy}$ (i.e. $I_0=1.1\sigma_{\text{\tiny rms}}$). The corresponding radio power at $1.4\,\text{GHz}$ is $P_\text{\tiny 1.4 GHz}=(1.4\pm0.6)\times10^{23}\,\text{W\,Hz}^{-1}$. This injected power can be considered as the approximate upper limit of the halo power and is consistent with our $1\sigma_{\text{\tiny rms}}$ estimate (i.e. $P_\text{\tiny 1.4 GHz}=(2.2\pm1.0)\times10^{23}\,\text{W\,Hz}^{-1}$ at 143 MHz) using the LOFAR low-resolution image. It is noted here that the halo emission is assumed to be distributed over an elliptical region between A1240-1 and A1240-2. In case that the $P_\text{\tiny 1.4 GHz}-R_\text{\tiny H}$ relation holds for A1240, the halo would be smaller (i.e. $a=b=R_\text{\tiny H}=264\,\text{kpc}$) and the upper limit for the halo power derived from simulations is $P_\text{\tiny 1.4 GHz}=(2.9\pm1.3)\times10^{22}\,\text{W\,Hz}^{-1}$. This is $\sim6$ times lower than the value derived from the VLA 1.4 GHz data in \cite{Bonafede2017} and is $\sim7$ times below the value predicted by the $P_\text{\tiny 1.4 GHz}-M_\text{\tiny 500}$ relation. The amount of flux that is recovered from the deconvolved images depends on the brightness of the injected halo model and the regions where the integrated flux is measured, as shown in Fig. \ref{fig:flux_recovered} for the case of the elliptical model for the halo.

The second possibility (ii) for the absence of the radio halo is beyond the scope of this paper as it requires mock haloes to be corrupted with realistic direction-independent and direction-dependent effects and injected into the $uv$-data prior to processing.

\section{Conclusions}
\label{sec:conclusions}

A1240 is a galaxy cluster in the process of merging \citep{Bonafede2009,Barrena2009}. Previous radio observations with the VLA (325 MHz and 1.4 GHz) by \cite{Bonafede2009} characterised two radio relics in the peripheral regions of the cluster. In this paper, we performed LOFAR 143 MHz observations of the cluster and combined these with archival GMRT 612 MHz and VLA 3 GHz data. With the wide  frequency range and deep observations, we provided new measurements on the spectral and polarimetric properties of the emission from the cluster relics and added further constraints on the existence of a radio halo. Below we summarize our results.

\begin{description}
	\item[$\bullet$] From the 143 MHz $-$ 3 GHz spectral index map we measured a spectral gradient across the width of the relics A1240-1 and A1240-2. The spectral indices steepen from $-0.94\pm0.05$ and $-0.97\pm0.05$ at the outer edges to  $-1.18\pm0.05$ and $-1.23\pm0.05$ at the inner edges for A1240-1 and A1240-2, respectively. Using the spectral index values at the locations of the shock fronts and assuming DSA, the corresponding Mach numbers for the shocks are $\mathcal{M}=2.4\pm{0.1}$ and $\mathcal{M}=2.3\pm{0.1}$, respectively. These Mach numbers are significantly lower than those (i.e. $\mathcal{M}_\text{\tiny A1240-1}=5.1^{+3.1}_{-1.1}$ and $\mathcal{M}_\text{\tiny A1240-2}=4.0^{+1.1}_{-0.6}$) estimated from the integrated spectral indices. 
	
	\item[$\bullet$] At $2-4$ GHz, we detected polarized emission in the regions of A1240-1 and A1240-2. The mean fractional polarization as measured from the VLA 480 MHz bandwidth data sets are $32\pm4\%$ and $17\pm4\%$ for A1240-1 and A1240-2, respectively. From these fractional polarization measurements we constrained the lower limits of the viewing angle of the merger axis to be $53\pm3^\circ$ and $39\pm5^\circ$ for A1240-1 and A1240-2, respectively.
	
	\item[$\bullet$]The magnetic field to thermal pressure ratio was estimated to be $\sim18\%$ and $\sim9\%$ in the regions of A1240-1 and A1240-2, respectively.
	
	\item[$\bullet$] The spectral gradients and polarimetric properties for the relics are consistent with the scenario that the relics were generated from shocks that are moving outwards. If the Mach numbers are 4.0 for A1240-1 and 5.1 for A1240-2, the required efficiency to accelerate electrons is less than $1\%$  and $0.5\%$, respectively, suggesting that the relativistic electrons in the relics might be accelerated directly from the thermal pool. If the relics have low Mach numbers (e.g. below 3), fossil electrons from radio galaxies in A1240 existing prior to the shock passage will be required. 
	
	\item[$\bullet$] With the sensitivity of our current LOFAR 143 MHz data, no hints of diffuse emission in the central regions of A1240 and A1237 were detected. Furthermore, no emission was observed in the region between the clusters. In the case that the A1240 halo spans the entire region (i.e. the ellipse in Fig. \ref{fig:lres}) between the relics, our upper limit for the radio halo power of A1240 by injecting of halo models is $P_\text{\tiny 1.4 GHz}=(1.4\pm0.6)\times10^{23}\,\text{W\,Hz}^{-1}$ which is consistent with the expected power from the $P_\text{\tiny 1.4 GHz}-M_\text{\tiny 500}$ scaling relation. However, in case of small halo (i.e. circle of radius $R_\text{\tiny H}=264\,\text{kpc}$) the estimated upper limit (i.e. $P_\text{\tiny 1.4 GHz}=(2.9\pm1.3)\times10^{22}\,\text{W\,Hz}^{-1}$) is $\sim7$ times below the $P_\text{\tiny 1.4 GHz}-M_\text{\tiny 500}$ relation. Our analysis suggests that we can reasonably exclude the presence of a small ($R_\text{\tiny H}=264\,\text{kpc}$) radio halo in A1240, while we cannot exclude the presence of a large elliptical radio halo extending between the two relics, since in this case our upper limit is consistent with the $P_\text{\tiny 1.4 GHz}-M_\text{\tiny 500}$ correlation.
	
\end{description}

\section*{Acknowledgements}

We thank the anonymous referees for the helpful comments. 
DNH, TS, RJvW and HR acknowledge support from the ERC Advanced Investigator programme NewClusters 321271. GB and RC acknowledge partial support from grant PRIN-INAF2014. AB acknowledges support from the ERC-Stg17 714245 DRANOEL. HA acknowledges the support of NWO via a Veni grant. SRON is supported financially by NWO, the Netherlands Organization for Scientific Research. GJW gratefully thanks the Leverhulme Trust for funding. This paper is based (in part) on results obtained with LOFAR equipment. LOFAR \citep{VanHaarlem2013} is the Low Frequency Array designed and constructed by ASTRON. We thank the staff of the GMRT that made these observations possible. GMRT is run by the National Centre for Radio Astrophysics of the Tata Institute of Fundamental Research. The National Radio Astronomy Observatory is a facility of the National Science Foundation operated under cooperative agreement by Associated Universities, Inc. The scientific results reported in this article are based in part on data obtained from the Chandra Data Archive, observations made by the Chandra X-ray Observatory and published previously in cited articles. This research has made use of software provided by the Chandra X-ray Center (CXC) in the application packages CIAO, ChIPS, and Sherpa. This work was performed under the auspices of the U.S. Department of Energy by Lawrence Livermore National Laboratory under Contract DE-AC52-07NA27344. We thank J. J. Harwood for discussions on the possibility of modelling of spectral ageing of radio relics. 



\bibliographystyle{mnras}


\noindent\textit{
	$^{1}$Leiden Observatory, Leiden University, PO Box 9513, NL-2300 RA Leiden, the Netherlands\\
	$^{2}$Netherlands Institute for Radio Astronomy (ASTRON), P.O. Box 2, 7990 AA Dwingeloo, The Netherlands\\
	$^{3}$Harvard-Smithsonian Center for Astrophysics, 60 Garden Street, Cambridge, MA 02138, USA\\
	$^{4}$SRON Netherlands Institute for Space Research, Sorbonnelaan 2, 3584 CA Utrecht, The Netherlands\\ 
	$^{5}$IRA INAF, via P. Gobetti 101 40129 Bologna, Italy\\
	$^{6}$Hamburger Sternwarte, University of Hamburg, Gojenbergsweg 112, 21029 Hamburg, Germany\\
	$^{7}$Lawrence Livermore National Lab, 7000 East Avenue, Livermore, CA 94550, USA\\
	$^{8}$Department of Physics, University of California, One Shields Avenue, Davis, CA 95616, USA\\
	$^{9}$SUPA, Institute for Astronomy, Royal Observatory, Blackford Hill, Edinburgh, EH9 3HJ, UK\\
	$^{10}$Dipartimento di Fisica e Astronomia, Universit\`a di Bologna, via P. Gobetti 93/2, 40129 Bologna, Italy\\ 
	$^{11}$Th\"uringer Landessternwarte, Sternwarte 5, 07778 Tautenburg, Germany\\
	$^{12}$European Southern Observatory, Karl-Schwarzschild-Str. 2, 85748, Garching, Germany\\
	$^{13}$Department of Physical Science, The Open University, Milton Keynes MK7 6AA, England\\
	$^{14}$Space Science Department, The Rutherford Appleton Laboratory, Chilton, Didcot, Oxon OX11 0NL, England\\ }


\bsp	
\label{lastpage}
\end{document}